\documentclass[12pt]{article}
\usepackage{cite}
\usepackage[dvips]{graphicx}
\usepackage{amsmath}
\usepackage{amssymb}

\begin{document}

\begin{flushright}
MAN/HEP/2010/22\\[-2pt]
\end{flushright}
\bigskip

\begin{center}
{\Huge {\bf Lepton Flavour Violation and {\boldmath $\theta_{13}$}\\[3mm]
in Minimal Resonant Leptogenesis}}\\[1.5cm] 
{\large Frank F. Deppisch$^{1,2}$\footnote{Email:
    f.deppisch@ucl.ac.uk} and Apostolos
  Pilaftsis$^1$\footnote{Email:
    apostolos.pilaftsis@manchester.ac.uk}}\\[0.3cm]  
{\em $^1$School of Physics and Astronomy, University of Manchester,}\\ 
{\em Manchester M13 9PL, United Kingdom}\\[0.1cm]
{\em $^2$Department of Physics and Astronomy, University College London,}\\ 
{\em London WC1E 6BT, United Kingdom}
\end{center}

\vspace{1.5cm} 

\centerline{\bf Abstract}
\noindent
{\small We  study the impact  of minimal non-supersymmetric  models of
  resonant leptogenesis  on charged  lepton flavour violation  and the
  neutrino  mixing  angle  $\theta_{13}$. Possible  low-scale  flavour
  realisations  of  resonant   $\tau$-,  $\mu$-  and  $e$-leptogenesis
  provide  very  distinct and  predictive  frameworks  to explain  the
  observed baryon asymmetry in the Universe by sphaleron conversion of
  an individual $\tau$-,  $\mu$- and $e$-lepton-number asymmetry which
  gets  resonantly enhanced  via out-of-equilibrium  decays  of nearly
  degenerate heavy  Majorana neutrinos.  Based  on approximate flavour
  symmetries,  we  construct  viable  scenarios of  resonant  $\tau$-,
  $\mu$- and  $e$-leptogenesis compatible with  universal right-handed
  neutrino masses at the  GUT scale, where the required heavy-neutrino
  mass  splittings  are  generated  radiatively.  The  heavy  Majorana
  neutrinos in  such scenarios  can be as  light as 100~GeV  and their
  couplings to two of the charged leptons may be large. In particular,
  we explicitly  demonstrate the compelling role that  the three heavy
  Majorana neutrinos play, in  order to obtain successful leptogenesis
  and  experimentally  testable  rates  for lepton  flavour  violating
  processes, such  as $\mu\to e\gamma$  and $\mu \to e$  conversion in
  nuclei.}


\newpage

\setcounter{equation}{0}
\section{Introduction}\label{sec:Introduction}

The observed baryon asymmetry in  the Universe (BAU), which amounts to
a baryon-to-photon ratio  of number densities $\eta_B\approx 6.2\times
10^{-10}$~\cite{WMAP,Komatsu:2010fb},  provides one  of  the strongest
pieces    of    evidence    for    physics   beyond    the    Standard
Model~(SM)~\cite{reviews}.   One interesting  scenario  for explaining
the BAU  is leptogenesis~\cite{FY}.  Leptogenesis does  have a profound
link to neutrinos and the  origin of their extraordinary small masses.
In particular, the famous seesaw mechanism~\cite{seesaw} can give rise
to  their small  observed masses  through the  presence  of superheavy
Majorana  neutrinos close  to the  Grand Unified  Theory  (GUT) scale,
$M_{\rm   GUT}~\approx~2\times10^{16}$~GeV.   These   GUT-scale  heavy
neutrinos, being singlets under the  SM gauge group, may possess large
Majorana masses that violate lepton  number~($L$) by two units.  In an
expanding  Friedmann--Lema\^itre--Robertson--Walker  (FLRW)  Universe,
the heavy Majorana neutrinos can  decay out of equilibrium and produce
a  net leptonic  asymmetry.  The  so-produced leptonic  asymmetry gets
rapidly  reprocessed into the  observed BAU~\cite{FY}  by equilibrated
$(B+L)$-violating sphaleron interactions~\cite{KRS}.

A potentially interesting alternative to GUT-scale leptogenesis is the
framework  of  low-scale  resonant  leptogenesis  (RL)~\cite{APRD,PU}.
Within  this  framework, the  lowering  of the  scale  may  rely on  a
dynamical    mechanism,    in    which   heavy-neutrino    self-energy
effects~\cite{LiuSegre}    on    the    leptonic   asymmetry    become
dominant~\cite{Paschos} and  get resonantly enhanced~\cite{APRD}, when
a pair of heavy Majorana neutrinos has a mass difference comparable to
the heavy neutrino decay widths.   As a consequence of thermal RL, the
heavy  Majorana mass  scale can  be as  low as  $\sim  100$~GeV, while
maintaining  agreement   with  the  solar   and  atmospheric  neutrino
data~\cite{PU}.   One  of  the advantages  of  low-scale RL  is  that  the
reheating  temperature $T_{\rm  reh}$ resulting  from  inflaton decays
does   not   need   to   be   very  high,   e.g.~$T_{\rm   reh}   \sim
1$--10~TeV~\cite{APreview},    thereby   avoiding    comfortably   the
overproduction of gravitinos in supersymmetric models whose  late decays may cause dissociation
of the light elements during the nucleosynthesis era~\cite{KL}.

Flavour  effects  due  to  heavy-neutrino  Yukawa  couplings  play  an
important  role  in  models  of RL~\cite{APtau,PU2}.   In  particular,
in~\cite{APtau}   a  scenario   was  put   forward,   called  resonant
$\tau$-leptogenesis~(R$\tau$L),  in which  the BAU  originates  from a
$\tau$-lepton  asymmetry, resonantly produced  by quasi-in-equilibrium
decays of heavy  Majorana neutrinos.  This mechanism makes  use of the
property  that  sphalerons  preserve  the individual  quantum  numbers
$\frac{1}{3}  B -  L_{e,\mu,\tau}$~\cite{KS,HT,DR,LS}.  In  a R$\tau$L
model, the generated  excess in the $L_\tau$ number  will be converted
into the  observed BAU, provided the  $L_\tau$-violating reactions are
not strong enough to wash out such an excess.  In such a scenario, the
heavy Majorana neutrinos can be  as light as 100~GeV and have sizeable
couplings to two of the charged leptons, specifically to electrons and
muons.    Consequently,   depending    on   the   flavour dynamics   of
heavy neutrino  Yukawa coupling  effects,  phenomenologically testable
models  of RL  can be  built that  could be  probed at  the LHC  or in
low energy   experiments    of   lepton number   violation~(LNV)   and
lepton flavour   violation~(LFV).    For   instance,  observables   of
particular     interest    are     the     neutrinoless    double-beta
($0\nu\beta\beta$) decay of heavy nuclei~\cite{CA}, the photonic decay
$\mu\to  e\gamma$  analyzed  by  the  MEG  experiment~\cite{MEG} and  the
coherent $\mu  \to e$  conversion in  nuclei to be  looked for  in the
planned  COMET/PRISM   experiment~\cite{PRISM}. 
% and  the  neutrinoless
%$\tau$-lepton decay $\tau\to\mu\mu\mu$ to be  searched for  by the
%LHCb collaboration~\cite{LHCb}.

In  this  paper  we  study  all  possible  alternatives  to  R$\tau$L,
including the minimal  models of resonant $\mu$-leptogenesis~(R$\mu$L)
and  resonant  $e$-leptogenesis~(R$e$L).   Collectively, we  refer  to
these three  different lepton-flavour realisations of  RL as R$\ell$L.
We   assume  that   the  R$\ell$L   models  have   an  SO(3)-symmetric
heavy neutrino mass spectrum at the GUT scale, with all heavy Majorana
neutrinos  being exactly  degenerate  at this  scale.   We consider  a
minimal non-supersymmetric framework, in which the heavy-neutrino mass
splittings    required    for     successful    RL    are    generated
radiatively~\cite{GJN}  and can therefore  be naturally  comparable to
the  decay widths of  the heavy  neutrinos.  Since  all charged-lepton
Yukawa  couplings  are  in  thermal  equilibrium  at  temperatures  $T
\stackrel{<}{{}_\sim}  10$~TeV~\cite{BCST},   we  consider  a  flavour
diagonal  basis for these  couplings, while  setting up  the Boltzmann
equations (BEs).  In  addition, we include the flavour  effects due to
individual heavy-neutrino Yukawa couplings~\cite{EMX,APtau}, which can
have  a  dramatic  impact  on  the  predictions  for  the  BAU  in  RL
models~\cite{APtau,PU2}.

The  layout  of  the  paper is  as  follows.   Section~\ref{sec:Model}
describes  the basic  structure of  the  minimal SM  with three  heavy
Majorana  neutrinos and  introduces the  flavour symmetries  needed to
realize  the   different  lepton-flavour  scenarios   associated  with
R$\ell$L.    As  mentioned   above,  we   assume  a   SO(3)  symmetric
heavy-neutrino   sector   at  the   GUT   scale   and  calculate   the
renormalization-group  (RG)  effects  on  the  mass  spectrum  of  the
electroweak-scale heavy Majorana  neutrinos and their Yukawa couplings
to charged  leptons.  Taking the light-neutrino  oscillation data into
account, we are  able to determine most of  the theoretical parameters
of the R$\ell$L models.  In Section~\ref{sec:LowEnergyObservables}, we
present analytic  results and predictions  for LFV observables  in the
three   different  R$\ell$L   models.   Section~\ref{sec:Leptogenesis}
briefly reviews the  basic framework of RL and  presents the BEs, upon
which our numerical  analysis is based. We also  clarify the necessity
of  having at least  three heavy  Majorana neutrinos  in RL  models in
order  to obtain  experimentally testable  LFV.  Section~\ref{sec:num}
presents  numerical estimates  of representative  R$\ell$L  models and
their  impact on  the neutrino  mixing angle  $\theta_{13}$.  Finally,
Section~\ref{sec:Conclusion} summarizes our conclusions.

\setcounter{equation}{0}
\section{Flavour Models of Minimal Resonant Leptogenesis}\label{sec:Model}

In  this   section,  we  describe  the   basic  theoretical  framework
underlying the different flavour  models of minimal RL. In particular,
our interest  is in  scenarios in  which the BAU  is generated  by the
production   of   an   individual  lepton   number~\cite{APtau}.   For
definiteness,  we  first consider  a  minimal  model  for R$\tau$L  in
Section~\ref{sec:RtauL}, and  then generalize  to the other  two cases
R$\mu$L  and  R$e$L   in  Sections~\ref{sec:RmuL}  and  \ref{sec:ReL},
respectively.

The  leptonic Yukawa  and  Majorana sectors  of  the SM  symmetrically
extended  with one  singlet right-handed  neutrino $\nu_{iR}$  per $i$
family (with $i=1,2,3 = e,\mu ,\tau$) are given by the Lagrangian
\begin{equation}
\label{Lym}
	-\; {\cal L}_{Y,M}\ =\
	 \bar{L} \Phi\,  {\bf h}^\ell\, l_{R}\ +\
	\bar{L}\tilde{\Phi}\, {\bf h}^{\nu}\, \nu_R\ +\ 
         \bar{\nu}_R^C\, {\bf m}_M\, \nu_R\ +\ \text{H.c.},
\end{equation}
where  $\Phi$ is  the SM  Higgs  doublet and  $\tilde{\Phi} =  i\tau_2
\Phi^*$  its  isospin conjugate.   Moreover,  we  have suppressed  the
generation index  $i$   from  the  left-handed  doublets  $L_i
=(\nu_{iL}, l_{iL})^T$, the  right-handed charged leptons $l_{iR}$ and
the right-handed  neutrinos $\nu_{iR}$, while  ordinary multiplication
between vectors and  matrices is implied~\footnote[1]{Occasionally, we
  will    also   denote   the    individual   lepton    numbers   with
  $L_{e,\mu,\tau}$,   but    hopefully   the   precise    meaning   of
  $L_{e,\mu,\tau}$ can be easily inferred from the context, without
  causing confusion.}.

To obtain  a phenomenologically relevant model in  this minimal setup,
at  least  3 singlet  heavy  Majorana  neutrinos $\nu_{1,2,3\,R}$  are
needed and  these have to  be nearly degenerate  in mass~\cite{APtau}.
To ensure  the latter,  we assume that  to leading order,  the singlet
Majorana sector is SO(3) symmetric, i.e.
\begin{equation}
\label{MSSO3}   
	{\bf m}_M\ =\ m_N {\bf 1}_3\ +\ {\bf \Delta m}_M\; , 
\end{equation}
where ${\bf 1}_3$ is the  $3\times 3$ identity matrix and ${\bf \Delta
  m}_M$ is a  general SO(3)-breaking matrix induced by  RG effects. As
we will discuss below,  compatibility with the observed light neutrino
masses  and   mixings  requires  that   $({\bf  \Delta  m}_M)_{ij}/m_N
\stackrel{<}{{}_\sim   }  10^{-7}$,   for   electroweak-mass  Majorana
neutrinos,  i.e.~for  $m_N \approx  0.1$--1~TeV.   We will  explicitly
demonstrate, how such an  SO(3)-breaking matrix ${\bf \Delta m}_M$, of
the required order, can be  generated radiatively via the RG evolution
of  the right-handed  neutrino mass  matrix ${\bf  m}_M$ from  the GUT
scale $M_X \approx 2\times 10^{16}~{\rm GeV}$ to the mass scale of the
right-handed neutrinos $m_N$.

To one-loop order, the RG equations governing the $3\times 3$ matrices
of  the neutrino  Yukawa couplings  ${\bf h}^\nu$,  the charged-lepton
Yukawa  couplings ${\bf  h}^\ell$  and the  singlet Majorana  neutrino
masses ${\bf m}_M$ are given by~\cite{Antusch}
\begin{eqnarray}
\label{eq:RGEhnu}
	\frac{d{\bf h}^\nu}{dt} &=& \frac{1}{16\pi^2}\;
	\bigg[\,
		\bigg(
			\frac{9}{4}g_2^2+\frac{3}{4}g_1^2-T
		\bigg)\mathbf{1}_3\
		-\ \frac{3}{2}\,
		\bigg(
			{\bf h}^\nu {\bf h}^{\nu\dagger} - {\bf
                          h}^\ell {\bf h}^{\ell\dagger}
		\bigg)\,
	\bigg]\, {\bf h}^\nu, \\[3mm]
\label{eq:RGEhell}
	\frac{d{\bf h}^\ell}{dt} &=& \frac{1}{16\pi^2}\;
	\bigg[\,
		\bigg(
			\frac{9}{4}g_2^2+\frac{15}{4}g_1^2-T
		\bigg)\mathbf{1}_3\
		+\  \frac{3}{2}\,
		\bigg(
			{\bf h}^\nu {\bf h}^{\nu\dagger} - {\bf
                          h}^\ell {\bf h}^{\ell\dagger}
		\bigg)\,
	\bigg] 
	{\bf h}^\ell, \\[3mm]
\label{eq:RGEM}
	\frac{d{\bf m}_M}{dt} &=& -\;\frac{1}{16\pi^2}\;
	\bigg[\,
		\big({\bf h}^{\nu\dagger} {\bf h}^\nu \big)\, {\bf m}_M\	+\ 
                {\bf m}_M \big({\bf h}^{\nu {\rm T}} {\bf h}^{\nu\,*}\big)\,
	\bigg]\; ,
\end{eqnarray}
where  $t  =  \ln(M_X/\mu)$  and  $\mu$ is  the  RG  evolution  scale.
Moreover, $g_1$ and $g_2$ are  the gauge couplings of the U(1)$_Y$ and
SU(2)$_L$  gauge  groups,  respectively,  and  $T$  is  the  shorthand
notation for the trace
\begin{equation} 
	T\ \equiv \ \text{Tr}
	\Big(
		3 {\bf h}^u {\bf h}^{u\dagger}\: +\: 
    3\, {\bf h}^d {\bf h}^{d\dagger}\: +\: {\bf h}^\nu {\bf h}^{\nu\dagger}\:
		 +\: {\bf h}^\ell {\bf h}^{\ell\dagger}
	\Big)\; .
\end{equation}
Here, ${\bf h}^u$ and ${\bf h}^d$ are the $3\times 3$ matrices for the
up- and  down-type Yukawa couplings,  respectively.  At the  GUT scale
$M_X$, we  impose the universal boundary condition:  ${\bf m}_M(M_X) =
m_N  {\bf 1}_3$. The  corresponding boundary  values for  the neutrino
Yukawa  couplings ${\bf  h}^\nu$  depend on  the  particular model  of
R$\ell$L, which we now discuss in detail.

\subsection{Resonant {\boldmath $\tau$}-Leptogenesis}\label{sec:RtauL}

In the physical charged-lepton  mass basis, the SO(3) symmetry imposed
on the singlet Majorana sector  at the GUT scale $M_X$ gets explicitly
broken by a set of neutrino Yukawa couplings to the subgroup of lepton
symmetries:  ${\rm U(1)}_{L_e+L_\mu}\times {\rm  U(1)}_{L_\tau}$.  The
flavour  charge assignments  that give  rise  to such  a breaking  are
presented in Table~\ref{tab:ChargesRtauL}.

\begin{table}[t]
\centering
\begin{tabular}{lccccc}
\hline
        & $L_e, e_R$ & $L_\mu, \mu_R$ & $L_\tau, \tau_R$ 
        & $\nu_1$ & $\nu_2\pm i\nu_3$    \\
\hline
$U(1)_{L_e+L_\mu}$ & +1 & +1 &  0 & 0 & $\pm 1$ \\
$U(1)_{L_\tau}$    &  0 &  0 & +1 & 0 &      0  \\
\hline
\end{tabular}
\caption{\it Flavour charge assignments for the breaking ${\rm SO(3)}\to
  {\rm U(1)}_{L_e+L_\mu}\times {\rm U(1)}_{L_\tau}$.}\label{tab:ChargesRtauL}  
\end{table}

As   a   consequence  of   the   ${\rm  U(1)}_{L_e+L_\mu}\times   {\rm
  U(1)}_{L_\tau}$ symmetry, the  neutrino Yukawa coupling matrix takes
on the general form:
\begin{equation}
 \label{hnutau}
	{\bf h}^\nu\ =\
	\left(\begin{array}{ccc}
		0  & a e^{-i\pi/4}  & a e^{i\pi/4} \\
		0  & b e^{-i\pi/4}  & b e^{i\pi/4} \\
		0  & 0                & 0 
	\end{array}\right)\ 
	+\ {\bf \delta h}^\nu\; ,
\end{equation}
and ${\bf \delta h}^\nu$ vanishes,  if the symmetry is exact.  In this
symmetric limit, the light neutrinos  remain massless to all orders in
perturbation theory, whilst $a$ and $b$ are free unconstrained complex
parameters. The phases accompanying these parameters in~(\ref{hnutau}) are simply chosen for convenience to maximize the lepton asymmetry in leptogenesis (see Section~\ref{sec:Asymmetries}) when $a$ and $b$ are real.  In order  to give  masses  to the  light neutrinos,  the
following  minimal departure  ${\bf  \delta h}^\nu$  from the  flavour
symmetric limit is considered:
\begin{equation}
  \label{dhnutau}
	{\bf \delta h}^\nu\ =\
\left(\begin{array}{ccc}
		\epsilon_e    & 0 & 0 \\
		\epsilon_\mu  & 0 & 0 \\
		\epsilon_\tau & \kappa_1 e^{-i(\pi/4-\gamma_1)} & 
		\kappa_2 e^{i(\pi/4-\gamma_2)}
	\end{array}\right)\; ,
\end{equation}
where  $|\epsilon_{e,\mu,\tau}|,\ \kappa_{1,2}  \ll  |a|,|b|$ and  the
phases $\gamma_{1,2}$ are  unrestricted.  A more precise determination
of the range  of parameter values can be  obtained from the low-energy
neutrino data and successful leptogenesis.

It  is important  to notice  that the flavour structure  of the  neutrino
Yukawa couplings ${\bf h}^\nu$  is preserved through the RG evolution,
as  long as  ${\bf \delta  h}^\nu$ remains  a small  perturbation.  In
detail, RG  effects violate the  SO(3)-invariant form of  the Majorana
mass  matrix ${\bf m}_M(M_X)  = m_N  {\bf 1}_3$  by the  3-by-3 matrix
${\bf   \Delta   m}_M$.   In   the   leading-log  approximation,   the
SO(3)-breaking matrix ${\bf \Delta m}_M$ reads:
\begin{eqnarray}
	\label{eq:mNFirstOrder}
	{\bf \Delta m}_M &=& 
	- \frac{m_N}{8\pi^2}\ln\left(\frac{M_X}{m_N}\right)
	\text{Re}\left[ {\bf h}^{\nu\dagger}(M_X) {\bf h}^\nu(M_X)
          \right]
        \ =\ - \frac{m_N}{8\pi^2}\ln\left(\frac{M_X}{m_N}\right)\\
        &&\hspace{-2.05cm}\times \begin{pmatrix}
		\epsilon_e^2 + \epsilon_\mu^2 +\epsilon_\tau^2  & 
		\frac{1}{\sqrt{2}}(a\epsilon_e+b\epsilon_\mu) 
	+ \epsilon_\tau\kappa_1 \sin\bar{\gamma}_1 &
        \frac{1}{\sqrt{2}}(a\epsilon_e+b\epsilon_\mu)
	+ \epsilon_\tau\kappa_2 \sin\bar{\gamma}_2 \\
    	\frac{1}{\sqrt{2}}(a\epsilon_e+b\epsilon_\mu) 
	+ \epsilon_\tau\kappa_1 \sin\bar{\gamma}_1 & 
		a^2+b^2+\kappa_1^2          &
		\kappa_1\kappa_2\sin (\gamma_1+\gamma_2) \\
	\frac{1}{\sqrt{2}}(a\epsilon_e+b\epsilon_\mu)
	+ \epsilon_\tau\kappa_2 \sin\bar{\gamma}_2 & 
		\kappa_1\kappa_2\sin(\gamma_1+\gamma_2) & 
		a^2+b^2+\kappa_2^2
	\end{pmatrix}\, ,\nonumber
\end{eqnarray}
with $\bar{\gamma}_{1,2} = \gamma_{1,2} + \frac{\pi}{4}$.
Correspondingly,  the  neutrino  and charged-lepton  Yukawa  couplings
modify via RG running from $M_X$ to $m_N$ as follows:
\begin{eqnarray}
  \label{eq:hnuFirstOrder}
	{\bf h}^\nu(m_N) &=&
	\left[\,
		\mathbf{1}_3\ 
		+ \
		\frac{\ln\left(\frac{M_X}{m_N}\right)}{16\pi^2}
		\begin{pmatrix}
				U-3a^2  & -3ab  & 0  \\
    			-3ab  & U-3b^2  & 0  \\
				0  & 0  & U+\frac{3}{2}h_\tau^2  \\
		\end{pmatrix}
	\right]\,
	{\bf h}^\nu(M_X)\; ,\\
\label{eq:mhellFirstOrder}
	{\bf h}^\ell(m_N) &=& \left[\,
		\mathbf{1}_3\ 
		+\
		\frac{\ln\left(\frac{M_X}{m_N}\right)}{16\pi^2}
		\begin{pmatrix}
			U'+3a^2  & 3ab  & 0  \\
   			3ab  & U'+3b^2  & 0  \\
				0  & 0  & U'-\frac{3}{2}h_\tau^2  \\
		\end{pmatrix}
	\right]\,
	{\bf h}^\ell (M_X)\; ,
\end{eqnarray}
where $U$ and $U'$ stand for the shorthand expressions 
\begin{eqnarray}
 U & \equiv &
    \frac{9}{4}g_2^2\: +\: \frac{3}{4}g_1^2\: -\: 3h_b^2\: -\: 
  3h_t^2\: -\: h_\tau^2\: -\: 2a^2\: -\: 2b^2\; ,\nonumber\\
 U' & \equiv & 
    \frac{9}{4}g_2^2\: +\: \frac{15}{4}g_1^2\: -\: 3h_b^2\: -\:
    3h_t^2\: -\: h_\tau^2\: -\: 2a^2\: -\: 2b^2\; .
\end{eqnarray}
Observe that the universal contributions $U$ and $U'$ are dominated by
the  top-quark  Yukawa coupling  $h_t$  and  are approximately  equal,
i.e.~$U \approx U'  \sim -\,3$.  For the models of  interest to us, we
have $a, b, h_\tau \sim 10^{-2}  \ll h_t$, so the RG effects give rise
to an overall rescaling of  the charged and neutrino Yukawa couplings,
${\bf h}^\ell$ and ${\bf h}^\nu$.   Without loss of generality, we may
assume that  the charged lepton Yukawa-coupling  matrix ${\bf h}^\ell$
is  positive  and diagonal  at  the  scale  $m_N$, i.e.~close  to  the
electroweak scale.   Given the form invariance of  ${\bf h}^\nu$ under
RG  effects, we  may  therefore  define all  input  parameters at  the
right-handed neutrino  mass scale $m_N$,  where the matching  with the
light neutrino data is performed.

We   may  now  determine   the  Yukawa   parameters  $(a,b,\epsilon_e,
\epsilon_\mu,  \epsilon_\tau)$, in  terms of  the  light-neutrino mass
matrix  ${\bf m}^\nu$  in  the positive  and  diagonal charged  lepton
Yukawa  basis.  To do  so, we  first notice  that the  chosen symmetry
${\rm  U(1)}_{L_e+L_\mu}\times {\rm  U(1)}_{L_\tau}$ is  sufficient to
ensure the vanishing of the  light neutrino mass matrix ${\bf m}^\nu$.
In fact, if it is an  exact symmetry of the theory, the light neutrino
mass   matrix   will   vanish    to   all   orders   in   perturbation
theory~\cite{AZPC}.   To   leading  order  in   the  symmetry-breaking
parameters ${\bf \Delta m}_M$  and $\delta{\bf h}^\nu$, the tree-level
light neutrino mass matrix ${\bf m}^\nu$ is given by
\begin{eqnarray}
  \label{mnutree}
	{\bf m}^\nu &=&
        -\ \frac{v^2}{2}\, {\bf h}^\nu\, {\bf m}_M^{-1} {\bf h}^{\nu
          {\rm T}}\  =\ \frac{v^2}{2m_N}\, 
	\bigg(\frac{{\bf h}^\nu {\bf \Delta m}_M {\bf h}^{\nu {\rm T}}}{m_N}\
		-\ {\bf h}^\nu {\bf h}^{\nu {\rm T}}
	\bigg) \nonumber\\
	&=&
	-\ \frac{v^2}{2m_N}
	\begin{pmatrix}
		\frac{\Delta m_N}{m_N}a^2-\epsilon_e^2  & 
		\frac{\Delta m_N}{m_N}ab-\epsilon_e\epsilon_\mu & 
		-\epsilon_e\epsilon_\tau \\
    	\frac{\Delta m_N}{m_N}ab-\epsilon_e\epsilon_\mu & 
		\frac{\Delta m_N}{m_N}b^2-\epsilon_\mu^2          &
		-\epsilon_\mu\epsilon_\tau \\
		-\epsilon_e\epsilon_\tau & 
		-\epsilon_\mu\epsilon_\tau &
		-\epsilon_\tau^2
	\end{pmatrix},
\end{eqnarray}
where $v = 2M_W/g_w = 245$~GeV  is the vacuum expectation value of the
SM    Higgs   field~$\Phi$.    In    deriving   the    last   equation
in~(\ref{mnutree}),  we have  also  assumed that  $\sqrt{\frac{\Delta
    m_N}{m_N}}\kappa_{1,2}\ll  \epsilon_{e,\mu,\tau}$,  where  $\Delta
m_N$ stands for the expression
\begin{eqnarray}
  \label{DmN}
	\Delta m_N &\equiv & 
   2({\bf \Delta m}_M)_{23}\: +\: i\Big[ ({\bf \Delta m}_M)_{33}
- ({\bf \Delta m}_ M)_{22} \Big]\nonumber\\
& = & -\ \frac{m_N}{8\pi^2}\ln\bigg(\frac{M_X}{m_N}\bigg)
	\Big[ 2\kappa_1\kappa_2\sin(\gamma_1+\gamma_2) 
		+ i(\kappa_2^2-\kappa_1^2) \Big]\; .
\end{eqnarray}
As a consequence of the flavour symmetry ${\rm U(1)}_{L_e+L_\mu}\times
{\rm     U(1)}_{L_\tau}$,     the    symmetry-violating     parameters
$\epsilon_{e,\mu,\tau}$ and $\kappa_{1,2}$  enter the tree-level light
neutrino mass matrix  ${\bf m}^\nu$ in (\ref{mnutree}) quadratically.
This in  turn implies that for electroweak-scale  heavy neutrinos $m_N
\sim v$,  the symmetry-breaking Yukawa couplings  ${\bf \delta h}^\nu$
in (\ref{dhnutau})  need not be  much smaller than the  electron Yukawa
coupling $h_e \sim 10^{-6}$.  Moreover, one should observe that only a
particular   combination  of   SO(3)-violating  terms   $({\bf  \Delta
  m}_M)_{ij}$   appears  in  ${\bf   m}^\nu$  through   $\Delta  m_N$.
Nevertheless,   for  electroweak-scale   heavy  neutrinos   with  mass
differences  $|\Delta  m_N|/m_N  \stackrel{<}{{}_\sim}  10^{-7}$,  one
should  have  $|a|,\,  |b|  \stackrel{<}{{}_\sim}  10^{-2}$  to  avoid
getting too large light neutrino masses much above~0.5~eV.

Given the  analytic form~(\ref{mnutree})  of the light  neutrino mass
matrix,  we  may  directly   compute  the  neutrino  Yukawa  couplings
$(a,b,\epsilon_e,  \epsilon_\mu,   \epsilon_\tau)$,  as  functions  of
$m_N$, the  phenomenologically constrained neutrino  mass matrix ${\bf
  m}^\nu$  and  the  symmetry-breaking parameters  $\kappa_{1,2}$  and
$\gamma_{1,2}$:
\begin{eqnarray}
\label{eq:a2}
	a^2 &=& 
	\frac{2m_N}{v^2}\ \frac{8\pi^2}{\ln(M_X/m_N)}\
	\bigg( m^\nu_{11} - \frac{(m^\nu_{13})^2}{m^\nu_{33}}\bigg)\
	\Big[ 2\kappa_1\kappa_2\sin(\gamma_1+\gamma_2) 
		+ i(\kappa_2^2-\kappa_1^2)\Big]^{-1}\; , \nonumber\\
	b^2 &=& 
	\frac{2m_N}{v^2}\ \frac{8\pi^2}{\ln(M_X/m_N)}\
	\bigg( m^\nu_{22} - \frac{(m^\nu_{23})^2}{m^\nu_{33}}\bigg)\
	\Big[ 2\kappa_1\kappa_2\sin(\gamma_1+\gamma_2) 
		+ i(\kappa_2^2-\kappa_1^2)\Big]^{-1}\; , \nonumber\\
	\epsilon_e^2 &=& 
	\frac{2m_N}{v^2}\ \frac{(m^\nu_{13})^2}{m^\nu_{33}}\; , \\
	\epsilon_\mu^2 &=& 
	\frac{2m_N}{v^2}\ \frac{(m^\nu_{23})^2}{m^\nu_{33}}\; , \nonumber\\
	\epsilon_\tau^2 &=& 
	\frac{2m_N}{v^2}\ m^\nu_{33}\; .\nonumber
\end{eqnarray}
Since  the  approximate   light-neutrino  mass  matrix  ${\bf  m}^\nu$
in~(\ref{mnutree})   has   rank   2,   the  lightest   neutrino   mass
eigenstate~$\nu_1$  will  be  massless  in  this  approximation.   The
relations  given in~(\ref{eq:a2})  will  be used  to obtain  numerical
estimates  of the  BAU, in  terms of  $m_N$ and  the symmetry-breaking
parameters  $\kappa_{1,2}$  and $\gamma_{1,2}$,  for  both normal  and
inverted hierarchy scenarios of light neutrinos.  In the following, we
discuss the two remaining flavour variants of RL: R$\mu$L and~R$e$L.

\subsection{Resonant {\boldmath $\mu$}-Leptogenesis}\label{sec:RmuL}

The  flavour scenario  of R$\mu$L  gets realized,  once  the GUT-scale
SO(3)  symmetry  gets broken  to  ${\rm U(1)}_{L_e+L_\tau}\times  {\rm
  U(1)}_{L_\mu}$.   The  flavour charge  assignments  related to  this
breaking   are  presented   in   Table~\ref{tab:ChargesRmuL}.   As   a
consequence, the  neutrino Yukawa coupling ${\bf h}^\nu$  takes on the
form:
\begin{equation}
 \label{hnumu}
	{\bf h}^\nu\ =\
	\left(\begin{array}{ccc}
		0  & a e^{-i\pi/4}  & a e^{i\pi/4} \\
		0  & 0                & 0 \\
	        0  & b e^{-i\pi/4}  & b e^{i\pi/4}
		\end{array}\right)\ 
	+\ {\bf \delta h}^\nu\; ,
\end{equation}
where the subdominant neutrino Yukawa-coupling matrix,
\begin{equation}
  \label{dhnumu}
	{\bf \delta h}^\nu\ =\
\left(\begin{array}{ccc}
\epsilon_e    & 0 & 0 \\
\epsilon_\mu  & \kappa_1 e^{-i(\pi/4-\gamma_1)} & \kappa_2 e^{i(\pi/4-\gamma_2)}\\
\epsilon_\tau & 0 & 0 
	\end{array}\right)\; ,
\end{equation}
breaks    minimally    the    ${\rm   U(1)}_{L_e+L_\tau}\times    {\rm
  U(1)}_{L_\mu}$ flavour symmetry.

\begin{table}[t]
\centering
\begin{tabular}{lccccc}
\hline
        & $L_e, e_R$ & $L_\mu, \mu_R$ & $L_\tau, \tau_R$ 
        & $\nu_1$ & $\nu_2\pm i\nu_3$ \\
\hline
$U(1)_{L_e+L_\tau}$ & +1 &  0 & +1 & 0 & $\pm 1$ \\
$U(1)_{L_\mu}$      &  0 & +1 &  0 & 0 &      0 \\
\hline
\end{tabular}
\caption{\it Flavour charge assignments for the breaking $SO(3)\to
  U(1)_{L_e+L_\tau}\times U(1)_{L_\mu}$.}\label{tab:ChargesRmuL} 
\end{table}

As in the R$\tau$L case, we assume that the Majorana-mass matrix ${\bf
  m}_M$ is proportional to~${\bf 1}_3$ at the GUT scale $M_X$ and gets
radiatively broken via RG effects at the heavy Majorana neutrino scale
$m_N$. Taking  into account both symmetry-breaking  terms ${\bf \Delta
  m}_M$  and  ${\bf \delta  h}^\nu$,  the  light-neutrino mass  matrix
acquires an analogous form in R$\mu$L:
\begin{eqnarray}
  \label{mnuRmuL}
	{\bf m}^\nu &=&
	-\ \frac{v^2}{2m_N}
	\begin{pmatrix}
		\frac{\Delta m_N}{m_N}a^2-\epsilon_e^2  & 
		-\epsilon_e\epsilon_\mu & 
		\frac{\Delta m_N}{m_N}ab -\epsilon_e\epsilon_\tau \\
    	-\epsilon_e\epsilon_\mu & 
		-\epsilon_\mu^2          &
		-\epsilon_\mu\epsilon_\tau \\
	\frac{\Delta m_N}{m_N}ab	-\epsilon_e\epsilon_\tau & 
		-\epsilon_\mu\epsilon_\tau &
	\frac{\Delta m_N}{m_N}b^2	-\epsilon_\tau^2
	\end{pmatrix}\; ,
\end{eqnarray}
where  $\Delta  m_N$ is  given  by~(\ref{DmN}).  Correspondingly,  the
neutrino  Yukawa coupling  parameters  $(a,b,\epsilon_e, \epsilon_\mu,
\epsilon_\tau)$  may  be  analogously  expressed, in  terms  of  ${\bf
  m}^\nu$, $m_N$, $\kappa_{1,2}$ and $\gamma_{1,2}$ as follows:
\begin{eqnarray}
\label{eq:a2RmuL}
	a^2 &=& 
	\frac{2m_N}{v^2}\ \frac{8\pi^2}{\ln(M_X/m_N)}\
	\bigg( m^\nu_{11} - \frac{(m^\nu_{12})^2}{m^\nu_{22}}\bigg)\
	\Big[ 2\kappa_1\kappa_2\sin(\gamma_1+\gamma_2) 
		+ i(\kappa_2^2-\kappa_1^2)\Big]^{-1}\; , \nonumber\\
	b^2 &=& 
	\frac{2m_N}{v^2}\ \frac{8\pi^2}{\ln(M_X/m_N)}\
	\bigg( m^\nu_{33} - \frac{(m^\nu_{23})^2}{m^\nu_{22}}\bigg)\
	\Big[ 2\kappa_1\kappa_2\sin(\gamma_1+\gamma_2) 
		+ i(\kappa_2^2-\kappa_1^2)\Big]^{-1}\; , \nonumber\\
	\epsilon_e^2 &=& 
	\frac{2m_N}{v^2}\ \frac{(m^\nu_{12})^2}{m^\nu_{22}}\; ,\\
	\epsilon_\mu^2 &=& 
	\frac{2m_N}{v^2}\ m^\nu_{22}\; ,\nonumber\\
        \epsilon_\tau^2 &=& 
	\frac{2m_N}{v^2}\ \frac{(m^\nu_{23})^2}{m^\nu_{22}}\ .\nonumber
\end{eqnarray}

\subsection{Resonant {\boldmath $e$}-Leptogenesis}\label{sec:ReL}

A third  possible flavour scenario  pertinent to R$e$L is  realized by
the    symmetry-breaking    pattern     ${\rm    SO}(3)    \to    {\rm
  U(1)}_{L_\mu+L_\tau}\times  {\rm  U(1)}_{L_e}$,  where  the  flavour
charge  assignments  are  given  in  Table~\ref{tab:ChargesReL}.   In
R$e$L, the neutrino Yukawa coupling ${\bf h}^\nu$ has the structure:
\begin{equation}
 \label{hnue}
	{\bf h}^\nu\ =\
	\left(\begin{array}{ccc}
		0  & 0                & 0 \\
   	        0  & a e^{-i\pi/4}  & a e^{i\pi/4} \\
		0  & b e^{-i\pi/4}  & b e^{i\pi/4}
		\end{array}\right)\ 
	+\ {\bf \delta h}^\nu\; ,
\end{equation}
and   the  breaking   terms  ${\bf   \delta  h}^\nu$   of   the  ${\rm
  U(1)}_{L_\mu+L_\tau}\times {\rm U(1)}_{L_e}$ flavour symmetry are
\begin{equation}
  \label{dhnue}
	{\bf \delta h}^\nu\ =\
\left(\begin{array}{ccc}
\epsilon_e    & \kappa_1 e^{-i(\pi/4-\gamma_1)} & \kappa_2 e^{i(\pi/4-\gamma_2)}\\
\epsilon_\mu  & 0 & 0 \\
\epsilon_\tau & 0 & 0 
	\end{array}\right)\; .
\end{equation}

\begin{table}[t]
\centering
\begin{tabular}{lccccc}
\hline
        & $L_e, e_R$ & $L_\mu, \mu_R$ & $L_\tau, \tau_R$ 
        & $\nu_1$ & $\nu_2\pm i\nu_3$ \\
\hline
$U(1)_{L_\mu+L_\tau}$ &  0 & +1 & +1 & 0 & $\pm 1$ \\
$U(1)_{L_e}$          & +1 &  0 &  0 & 0 &      0  \\
\hline
\end{tabular}
\caption{\it Flavour charge assignments for the breaking $SO(3)\to
  U(1)_{L_\mu+L_\tau}\times U(1)_{L_e}$.}\label{tab:ChargesReL} 
\end{table}

In close analogy with the previous two scenarios of R$\tau$L and R$\mu$L,
the light neutrino mass matrix in R$e$L is given by 
\begin{eqnarray}
  \label{mnuReL}
	{\bf m}^\nu &=& -\ \frac{v^2}{2m_N}
	\begin{pmatrix}
		-\epsilon_e^2  & 
		-\epsilon_e\epsilon_\mu & 
		 -\epsilon_e\epsilon_\tau \\
    	-\epsilon_e\epsilon_\mu & 
		\frac{\Delta m_N}{m_N}a^2 -\epsilon_\mu^2          &
		\frac{\Delta m_N}{m_N}ab -\epsilon_\mu\epsilon_\tau \\
		-\epsilon_e\epsilon_\tau & 
		\frac{\Delta m_N}{m_N}ab -\epsilon_\mu\epsilon_\tau &
	\frac{\Delta m_N}{m_N}b^2	-\epsilon_\tau^2
	\end{pmatrix},
\end{eqnarray}
where  $\Delta  m_N$ retains  its  analytic  form of~(\ref{DmN}).   By
analogy, we may  derive in R$e$L the relations  of the neutrino Yukawa
coupling parameters $(a,b,\epsilon_e, \epsilon_\mu, \epsilon_\tau)$ to
${\bf  m}^\nu$, $m_N$,  $\kappa_{1,2}$ and~$\gamma_{1,2}$.  These are
given by
\begin{eqnarray}
\label{eq:a2ReL}
	a^2 &=& 
	\frac{2m_N}{v^2}\ \frac{8\pi^2}{\ln(M_X/m_N)}\
	\bigg( m^\nu_{22} - \frac{(m^\nu_{12})^2}{m^\nu_{11}}\bigg)\
	\Big[ 2\kappa_1\kappa_2\sin(\gamma_1+\gamma_2) 
		+ i(\kappa_2^2-\kappa_1^2)\Big]^{-1}\; , \nonumber\\
	b^2 &=& 
	\frac{2m_N}{v^2}\ \frac{8\pi^2}{\ln(M_X/m_N)}\
	\bigg( m^\nu_{33} - \frac{(m^\nu_{13})^2}{m^\nu_{11}}\bigg)\
	\Big[ 2\kappa_1\kappa_2\sin(\gamma_1+\gamma_2) 
		+ i(\kappa_2^2-\kappa_1^2)\Big]^{-1}\; , \nonumber\\
	\epsilon_e^2 &=& 
	\frac{2m_N}{v^2}\ m^\nu_{11}\;  , \\
	\epsilon_\mu^2 &=& 
	\frac{2m_N}{v^2}\ \frac{(m^\nu_{12})^2}{m^\nu_{11}}\; , \nonumber\\
	 \epsilon_\tau^2 &=& 
	\frac{2m_N}{v^2}\ \frac{(m^\nu_{13})^2}{m^\nu_{11}}\ .\nonumber
\end{eqnarray}

\setcounter{equation}{0}
\section{Low Energy Observables}\label{sec:LowEnergyObservables}

Low-energy neutrino  data provide  indisputable evidence for
neutrino oscillations.   We use these data to  determine and constrain
the fundamental  parameters of  the theory. We  also present  the full
range of  predictions for  the half life  of neutrinoless  double beta
decay of  a heavy nucleus,  within the different flavour  scenarios of
RL. In the same context, we present analytic results and estimates for
LFV observables, such as $\mu  \to e\gamma$ and $\mu \to e$ conversion
in nuclei.

\subsection{Light Neutrino Oscillation Data}\label{sec:OscillationData}

The  interpretation of  the  experimental results  on solar  neutrinos
suggests  that  $\nu_e  \to  \nu_{\mu,\tau}$ oscillations  are  mainly
driven by the mass squared difference $\Delta m^2_{12} = m_{\nu_2}^2 -
m_{\nu_1}^2$, while the corresponding experimental data on atmospheric
neutrinos are  interpreted by $\nu_{\mu  }\to \nu_{\tau}$ oscillations
driven  by $\Delta  m^2_{23} =  m_{\nu_3}^2 -  m_{\nu_2}^2$,  within a
minimal scheme  of three active neutrinos.  For  the present analysis,
we use the global fits  performed in \cite{JVdata}, even though global
fits of other groups  give compatible results~\cite{Concha}.  The best
fit values for the light  neutrino masses and mixings, including their
uncertainties at the $2\sigma$ level, are given by
\begin{eqnarray}
	\label{eq:LightNeutrinoBenchmark}
	\sin^2\theta_{12} \! &=&\! 0.32^{+0.05}_{-0.04}\;, \qquad
	\sin^2\theta_{23}\ =\ 0.50^{+0.13}_{-0.12}\; , \qquad
	\sin^2\theta_{13}\ =\ 0.007^{+0.026}_{-0.007}\;, \nonumber\\
	\Delta m_{12}^2 \! &=&\! \left(7.6^{+0.5}_{-0.3}\right)\times
        10^{-5}\text{ eV}^2\; , \qquad 
	|\Delta m_{13}^2| \ =\  \left(2.4^{+0.3}_{-0.3}\right)\times
        10^{-3}\text{ eV}^2\; , 
\end{eqnarray}
where the  sign of  $\Delta m_{13}^2 \equiv  \Delta m_{12}^2  + \Delta
m_{23}^2$ remains  still undetermined,  corresponding to the  cases of
the  so-called  normal  ($\Delta  m_{13}^2>0$) and  inverted  ($\Delta
m_{13}^2<0$)  neutrino mass hierarchy,  respectively.  As  outlined in
Section~\ref{sec:Model}, we use the oscillation parameters as input to
reduce the number  of free parameters in the  neutrino Yukawa coupling
matrix ${\bf h}^\nu$.

\subsection{{\boldmath $0\nu\beta\beta$} Decay}\label{sec:0nubb}

Models that  include Majorana neutrinos violate the  $L$-number and so
can give rise to  neutrinoless double beta decay ($0\nu\beta\beta$) of
a    heavy   nucleus   $^{A}_{Z}X$,    where   two    single   $\beta$
decays~\cite{doi85,Klapdor,HKK}   can  occur  simultaneously   in  one
nucleus, $^{A}_{Z}X  \to ^A_{Z+2}X +  2 e^-$.  The measurement  of the
half life of this decay  provides further information on the structure
of  the  light neutrino  mass  matrix  ${\bf  m}^\nu$. The  half  life
$T^{0\nu\beta\beta}_{1/2}$  for a  $0\nu\beta\beta$ decay  mediated by
light Majorana neutrinos is given by
\begin{equation}
\label{t1/2}
	\left(T^{0\nu\beta\beta}_{1/2}\right)^{-1}\ =\ \frac{\langle m
	\rangle^2}{m^2_e} |{\cal M}_{0\nu\beta\beta}|^2 G_{01}\; ,
\end{equation} 
where  $\langle m  \rangle$  denotes the  effective Majorana  neutrino
mass, $m_e$  is the electron mass and  ${\cal M}_{0\nu\beta\beta}$ and
$G_{01}$ are the nuclear matrix element and the phase space factor of the decay,
respectively. The effective neutrino  mass $\langle m\rangle$ is given
by the entry $\{11\}\ \equiv \{ee\}$ of the light neutrino mass matrix
${\bf m}^\nu$, which can be expressed as
\begin{equation}
	\langle m \rangle\ \equiv\ |{\bf m}^\nu_{ee}|\
	=\ \bigg|\sum_{i=1}^3 (U^\nu_{ei})^2\; m_{\nu_i}\bigg|\; , 
\end{equation}
where  $U^\nu$  is the  PMNS  neutrino-mixing matrix~\cite{PMNS}.   As
described in Section~\ref{sec:Model},  R$\ell$L models realise a light
neutrino  mass  spectrum  with  a vanishing  lightest  neutrino  mass,
$m_{\nu_1}=0$,  and either a  normal or  inverted mass  hierarchy. The
prediction  for the  effective Majorana  neutrino mass  for  these two
scenarios is given by
\begin{equation}
\label{eq:m0vbbnormalinv}
	\langle m \rangle\ =\ 
	\begin{cases}
		(2.5^{+3.0}_{-2.5})\times 10^{-3} & \Delta m_{13}^2>0 \\
		(2.9^{+3.0}_{-1.9})\times 10^{-2} & \Delta m_{13}^2<0
	\end{cases}\ .
\end{equation}
The  above  prediction  takes  into  account the  uncertainty  of  the
observed  oscillation  parameters  at  the  $2\sigma$  level  and  the
variation of the unknown Dirac and Majorana phases.  These predictions
are to be compared to the experimental bound $T_{1/2}^{0\nu\beta\beta}
> 1.9 \times  10^{25}$~years in the isotope $^{76}$Ge  reported by the
Heidelberg--Moscow collaboration~\cite{HMexp}, implying an upper limit
on $\langle m \rangle$ in the range:
\begin{equation}
	\langle m \rangle_{\rm exp}\ <\ (0.3 - 0.6)\text{ eV}\; .
\end{equation}
Here, the main uncertainty is due to the choice for the nuclear matrix
element  that occurs  in  (\ref{t1/2}). Future  $0\nu\beta\beta$-decay
experiments are expected to probe $\langle m \rangle$ to sensitivities
of order  $10^{-2}$~\cite{CA} and  so fall within  the range  given in
(\ref{eq:m0vbbnormalinv})  to validate the  mass scenario  of inverted
hierarchy.

\subsection{\boldmath$l_1\to l_2\gamma$}\label{sec:muegamma}

Heavy Majorana-neutrino loop effects may induce sizeable LFV couplings
to the  photon and  the $Z$  boson. These couplings  give rise  to LFV
decays, such  as $\mu \to e\gamma$~\cite{CL},  $\mu \to eee$~\cite{IP}
and  $\mu  \to  e$ conversion  in  nuclei.   The  strength of  LFV  is
controlled by the effective coupling matrix
\begin{equation}
\label{eq:Omega}
	{\bf \Omega}_{l_1 l_2}\ =\ 
	\frac{v^2}{2m_N^2}\ ({\bf h}^\nu {\bf h}^{\nu\dagger})_{l_1 l_2}\; ,
\end{equation}
which  governs  the flavour  transition  between  the charged  leptons
$l_{1,2}  =  e,\mu,\tau$  in  LFV  processes. The  LFV  decay  $l_1\to
l_2\gamma$, $l_{1}\in \{\mu,\tau\}$, with $l_2 \in\{\mu,e\}$, $l_2\neq
l_1$, whose branching fraction is given by
\begin{equation}
\label{Bllgamma}
	B(l_1\to l_2\gamma )\ =\ 
	\frac{\alpha^3_w s^2_w}{256\,\pi^2}\:
	\frac{m^4_{l_1}}{M^4_W}\:
	\frac{m_{l_1}}{\Gamma_{l_1}}\: |G_\gamma^{l_1 l_2}|^2\; .
\end{equation}
In the  above, $\Gamma_{l_1}$ is the  decay width of  lepton $l_1$ and
$G_\gamma^{\mu e}$ is a composite form-factor given by~\cite{IP}
\begin{equation}
	\label{eq:Ggammall}
	G_\gamma^{l_1 l_2}\ =\ 
	-{\bf\Omega}_{l_1 l_2}
        G_\gamma\left(\frac{m_N^2}{m_W^2}\right)\; ,
\end{equation}
with
\begin{equation}
	\label{eq:Ggamma}
	G_\gamma(x)\ =\ 
	-\ \frac{2x^3+5x^2-x}{4(1-x)^3}-\frac{3x^3}{2(1-x)^4}\ln x\; .
\end{equation}
Given that the  experimentally measured muon and tau  decay widths are
$\Gamma_\mu = 2.997\times 10^{-19}$~GeV and $\Gamma_\tau = 2.158\times
10^{-12}$~GeV~\cite{PDG}, the LFV branching ratios can be expressed as
\begin{align}
\label{eq:BllgammaApprox}
	B(\mu\to e\gamma)\ &\approx \
	8.0\cdot 10^{-4}\times  g\left(\frac{m_N}{m_W}\right)\; 
        |{\bf\Omega}_{\mu e}|^2\; ,\\
	B(\tau\to l_2\gamma)\ &\approx \
	1.5\cdot 10^{-4}\times g\left(\frac{m_N}{m_W}\right)\; 
        |{\bf\Omega}_{\tau l_2}|^2\; ,
	\quad l_2=e,\mu\; .
\end{align}
Here,  we defined $g(x)\equiv  4G^2_\gamma(x^2)$, which  possesses the
limits $g\to 1$  for $m_N \gg m_W$ and $g\to  1/16$ for $m_N=m_W$. Our
theoretical predictions will be contrasted to the current experimental
upper limits~\cite{PDG}
\begin{align}
\label{Bexpllgamma}
	B_{\rm exp} (\mu\to e\gamma)\   &<\ 1.2 \times 10^{-11}\; , \nonumber\\
	B_{\rm exp} (\tau\to\mu\gamma)\ &<\ 6.8 \times 10^{-8}\; , \\
	B_{\rm exp} (\tau\to e\gamma)\  &<\ 1.1 \times 10^{-7}\; , \nonumber
\end{align}
and the expected sensitivity of the MEG experiment\cite{MEG},
\begin{equation}
\label{BexpllgammaMEG}
	B_{\rm MEG} (\mu\to e\gamma)\ \approx\ 10^{-13}\; .
\end{equation}
In R$\ell$L models, only two of the right-handed neutrinos, $\nu_{2R}$
and $\nu_{3R}$,  have appreciable $e$-  and $\mu$-Yukawa couplings,
$a,b\approx  10^{-2}$,  $\kappa_{1,2}=10^{-5}-10^{-3}$,  and  will  be
relevant  to LFV  effects.  For  example, in  the R$\tau$L  model, the
parameters $|{\bf\Omega}_{l_1  l_2}|^2$ are, to  a good approximation,
given by
\begin{equation}
	|{\bf\Omega}_{\mu e}|^2\ \approx\ \frac{v^4}{m_N^4}a^2 b^2\; , \quad
	|{\bf\Omega}_{\tau\mu}|^2\ \approx\ 
	\frac{v^4}{m_N^4}\text{max}(\kappa^2_1,\kappa^2_2) b^2\; , \quad
	|{\bf \Omega}_{\tau e}|^2\ \approx\ 
	\frac{v^4}{m_N^4}\text{max}(\kappa^2_1,\kappa^2_2) a^2\; .
\end{equation}
Because of the relations  (\ref{eq:a2}), $a$ and $b$ are approximately
inversely   proportional   to   $\kappa_{1,2}$,  $a,b\approx   3\times
10^{-7}\kappa_{1,2}^{-1}$ (in the case of a normal light  neutrino mass hierarchy), and
for a typical value of  $\kappa_{1,2}\approx 10^{-4}$, $a$ and $b$ are
of  the order  $a,b\approx  3\times 10^{-3}$.  Thus,  in the  R$\tau$L
scenario, the following typical values for the branching ratios of the
photonic LFV $\mu$ and $\tau$ decays are predicted:
\begin{equation}
	B(\mu\to e\gamma)_{\rm R\tau L}\ \approx\ 10^{-13}\; , \qquad
	B(\tau\to l_2\gamma)_{\rm R\tau L}\ \approx\ 10^{-17}\; .
\end{equation}
Consequently, only $B(\mu\to e\gamma)$  is expected to be within reach
of future  experiments, whereas the  LFV $\tau$ decays are  far beyond
the realm of detection.  By analogy, the predictions in the R$e(\mu)$L
model are found to be
\begin{equation}
 B(\mu\to e\gamma)_{{\rm R} e(\mu) {\rm L}}\ \approx\ 10^{-16}\;, \qquad
 B(\tau\to l_2\gamma)_{{\rm R} e (\mu) {\rm L}}\ \approx\
 10^{-14}\; .
\end{equation}
Again, $B(\tau\to\mu\gamma)$ and  $B(\tau\to e\gamma)$ are expected to
be below  current and future experimental  sensitivities for parameter
choices compatible with  successful leptogenesis. While the prediction
for $B(\mu\to e\gamma)$ is far below the expected MEG sensitivity, the
LFV  $\mu \to  e$ transition  rate might  still be  high enough  to be
testable  at future experiments  measuring $\mu  \to e$  conversion in
nuclei  (see next  section).  Because  of the  inverse proportionality
between  $a,b$   and  $\kappa_{1,2}$  the   prediction  for  $B(\mu\to
e\gamma)$ is essentially independent of the choice for $\kappa_{1,2}$.
In  Section~\ref{sec:num}, we present  detailed numerical  results for
LFV $\mu$- and $\tau$-decays that confirm the validity of these simple
estimates.

\subsection{Coherent {\boldmath $\mu \to e$} Conversion in Nuclei}

One of the most sensitive probes  of LFV is the coherent conversion of
$\mu \to e$ in nuclei~\cite{FW,JDV}.   The $\mu \to e$ conversion rate
in    a   nucleus    with   nucleon    numbers   $(N,Z)$    is   given
by~\cite{FW,JDV,IP2}
\begin{equation}
\label{Bmueconv}
	B_{\mu e} (N,Z)\ \equiv\ 
	\frac{\Gamma [\mu\, (N,Z)\to e\, (N,Z) ] }{
	\Gamma [ \mu\, (N,Z) \to {\rm capture}]}\
	\approx\ \frac{ \alpha^3_{\rm em} \alpha^4_w m^5_\mu}{16\pi^2
	m^4_W \Gamma_{\rm capt} }\, \frac{Z^4_{\rm eff}}{Z}\,
	|F(-m^2_\mu)|^2\, |Q_W|^2\, ,
\end{equation}
where $\alpha_{\rm em} =  1/137$ is the electromagnetic fine structure
constant,   $Z_{\rm  eff}$   is  the   effective  atomic   number  and
$\Gamma_{\rm   capt}$  is   the   muon  nuclear   capture  rate.   For
${}^{48}_{22}{\rm  Ti}$, experimental  measurements give  $Z_{\rm eff}
\approx  17.6$~\cite{Zeff} and  $\Gamma_\text{capt}            \approx            1.705\times
10^{-18}$~GeV~\cite{SMR}.  Moreover, $|F(-m^2_\mu)|  \approx  0.54$ is
the nuclear form factor \cite{FPap} and
\begin{equation}
	Q_W\ =\ V_u (2Z +N)\: +\: V_d (Z+2N)
\end{equation}
is the weak matrix element, where
\begin{align}
\label{Vu}
	V_u\ &=\ 
	- 
	\left[
		1 + \frac{1}{3}s^2_w +
		\left(\frac{3}{8} - \frac{8}{9}s^2_w\right)
		\ln\frac{m_N^2}{m_W^2}
	\right]{\bf \Omega}_{\mu e}\: +\:
	\frac{1}{2}
	\left(\frac{1}{4}+\frac{2}{3}s^2_w\right)
	\frac{m_N^2}{m_W^2}({\bf\Omega}^2)_{\mu e}\; , \\
	\label{Vd}
	V_d\ &=\ \quad
	\left[
		\frac{1}{4} - \frac{1}{6}s^2_w + 
		\left(\frac{3}{8} - \frac{4}{9}s^2_w\right)
		\ln\frac{m_N^2}{m_W^2}
	\right] {\bf\Omega}_{\mu e}\:
	+\: \frac{1}{2}\left(\frac{1}{4}-\frac{1}{3}s^2_w \right)
	\frac{m_N^2}{m_W^2}({\bf\Omega}^2)_{\mu e}\; .
\end{align}
In the  case of  ${}^{48}_{22}{\rm Ti}$, $B_{\mu  e} (26,22)$  is then
approximately related to $B(\mu\to e\gamma)$ through
\begin{equation}
  \label{Rmue}
	B_{\mu e}(26,22)\ \approx\ 10^{-1} \times B(\mu\to e\gamma),
\end{equation}
for a right-handed neutrino mass  scale $m_N \approx 100$~GeV.  On the
experimental side, the strongest upper  bound is obtained from data on
$\mu\to e$ conversion in ${}^{48}_{22}{\rm Ti}$ \cite{SINDRUM},
\begin{equation}
\label{mueconvexp}
	B^{\rm exp}_{\mu e} (26,22)\ <\ 4.3 \times 10^{-12}.
\end{equation}
Using the  relation~(\ref{Rmue}), we observe that  this sensitivity is
comparable to  the current bound on $B(\mu\to  e\gamma)$. However, the
proposed COMET  and mu2e experiments, measuring  $\mu\to e$ conversion
in ${}^{27}_{13}{\rm Al}$, are  expected to be sensitive to conversion
rates of order $10^{-16}$ \cite{Kurup:2010zz,Glenzinski:2010zz}, thereby improving the sensitivity compared
to the current limit by four orders of magnitude.

\setcounter{equation}{0}
\section{Leptogenesis}\label{sec:Leptogenesis}

In  this  section,  we  briefly  review the  central  results  of  the
field-theoretic  formalism for  RL  developed in~\cite{APRD,PU}  which
will be  used in  our analysis.  We  then set  up the BEs  and present
approximate  solutions  for  the   kinematic  regime  of  interest  to
us. Finally,  we clarify the  necessity of introducing at  least three
right-handed neutrinos  into the theory,  in order to explain  the BAU
and obtain  testable rates  of LFV, such  as $B(\mu \to  e\gamma) \sim
10^{-12}$--$10^{-13}$.

\subsection{Leptonic Asymmetries}\label{sec:Asymmetries}

Within  the  framework  of   leptogenesis,  a  net  non-zero  leptonic
asymmetry results  from the CP-violating decays of  the heavy Majorana
neutrinos $N_\alpha$ into the  left-handed charged leptons $l^-_L$ and
light neutrinos  $\nu_{lL}$.  Consequently,  we have to  calculate the
partial decay width  of the heavy Majorana neutrino  $N_\alpha$ into a
particular lepton flavour $l$,
\begin{equation}
\label{GammaN}
	\Gamma_{\alpha l}\ =\
	\Gamma(N_\alpha \to l^-_L + W^+)\: +\: 
	\Gamma(N_\alpha\to \nu_{lL} + Z, H)\; . 
\end{equation}
For temperatures  above the electroweak  phase transition, the  SM VEV
vanishes  and  only  the  would-be  Goldstone  and  Higgs  modes  will
predominantly  contribute  to  $\Gamma_{\alpha  l}$.   

In   RL  models,  resumming   the  absorptive   parts  of   the  heavy
Majorana-neutrino self-energy transitions $N_\beta \to N_\alpha$ plays
an   important   role    in   the   computation   of   $\Gamma_{\alpha
l}$~\cite{APRD,PU}.   In order to  take this  resummation consistently
into  account,   we  first  introduce   the  lepton-flavour  dependent
absorptive transition amplitude~\cite{Pilaftsis:2008qt}
\begin{equation}
  \label{eq:Absl}
	A_{\alpha\beta}^l({\bf h}^\nu)\ \equiv\
	\frac{{\bf h}^{\nu}_{l\alpha} {\bf
            h}^{\nu*}_{l\beta}}{16\pi}\ ,
\end{equation}
which represents the contribution of a single charged lepton and light
neutrino flavour~$l$  running in the loop.  Summing  over all flavours
$l$, we then get the total transition amplitude
\begin{equation}
  \label{eq:Abs}
	A_{\alpha\beta}({\bf h}^\nu)\ \equiv\ 
	\sum_{l=e,\mu,\tau} A_{\alpha\beta}^l ({\bf h}^\nu)\ =\
	\frac{({\bf h}^{\nu\dagger} {\bf h}^\nu)^*_{\alpha\beta}}{16\pi}\ .
\end{equation}
Note  that  the diagonal  transition  amplitude $A_{\alpha\alpha}$  is
related to the  tree-level decay width of the  heavy Majorana neutrino
$N_\alpha$    through:     $\Gamma^{(0)}_{N_\alpha}    =    2\,m_{N_\alpha}
A_{\alpha\alpha}({\bf h^\nu})$.

Since  all   charged-lepton  Yukawa  couplings  will   be  in  thermal
equilibrium in  the low-scale leptogenesis scenarios  of our interest,
we consider  the weak basis in  which the matrices  ${\bf h}^\ell$ and
${\bf  m}_M$   are  both  diagonal  and  positive.    To  account  for
unstable-particle  mixing   effects  between  the   3  heavy  Majorana
neutrinos, we follow~\cite{APRD,PU}  and define the resummed effective
Yukawa   couplings    $\overline{\bf   h}^\nu_{l\alpha}$   and   their
CP-conjugate ones $\overline{\bf  h}^{\nu C}_{l\alpha}$ related to the
vertices $L\tilde{\Phi}  N_\alpha$ and $L^C  \tilde{\Phi}^* N_\alpha$,
respectively.  The  resummed neutrino Yukawa  couplings $\overline{\bf
  h}^\nu_{l\alpha}$ are given by~\cite{PU,Pilaftsis:2008qt}
\begin{eqnarray}
  \label{hres3g}
	\overline{\bf h}^\nu_{l\alpha} ({\bf h}^\nu) \! &=&\! 
        {\bf h}^\nu_{l\alpha}\: -\: i \sum\limits_{\beta,\gamma = 1}^3 
	|\varepsilon_{\alpha\beta\gamma}|\: {\bf h}^\nu_{l\beta} \\
	&&\hspace{-2.2cm}\times\
	\frac{
		m_\alpha ( m_\alpha A_{\alpha\beta} +
		m_\beta A_{\beta\alpha}) - i R_{\alpha \gamma} \left[ m_\alpha
		A_{\gamma\beta} ( m_\alpha A_{\alpha\gamma} + 
		m_\gamma A_{\gamma\alpha}) + 
		m_\beta A_{\beta\gamma} ( m_\alpha A_{\gamma\alpha} + 
		m_\gamma A_{\alpha \gamma}) \right]
	}{ 
		m^2_\alpha - m^2_\beta +
		2i m^2_\alpha A_{\beta\beta} + 2i{\rm Im}R_{\alpha\gamma} 
		\left( m^2_\alpha |A_{\beta\gamma}|^2 + 
		m_\beta m_\gamma {\rm Re}A^2_{\beta\gamma}\right)
	}\ ,\nonumber
\end{eqnarray}
where $|\varepsilon_{\alpha\beta\gamma}|$ is  the modulus of the usual
Levi--Civita     anti-symmetric     tensor    ($\varepsilon_{123}=1$),
$m^2_\alpha    \equiv m^2_{N_\alpha}   $,   $A_{\alpha\beta}    \equiv
A_{\alpha\beta}({\bf h}^\nu)$ and
\begin{equation}
	\label{eq:Rab}
	R_{\alpha \beta}\ =\ \frac{m^2_\alpha}{m^2_\alpha - m^2_\beta 
	+ 2i m^2_\alpha A_{\beta\beta}}\ .
\end{equation}
The respective CP-conjugate  effective Yukawa couplings $\overline{\bf
  h}^{\nu C}_{l\alpha}({\bf  h}^\nu)$ are obtained from~(\ref{hres3g})
by  replacing the  tree-level couplings  ${\bf  h}^\nu_{l\alpha}$ with
their complex conjugates ${\bf h}^{\nu *}_{l\alpha}$:
\begin{equation}
\label{hres3gc}
	\overline{\bf h}^{\nu C}_{l\alpha} ({\bf h}^\nu)\ =\  
	\overline{\bf h}^\nu_{l\alpha}({\bf h}^{\nu*})\; .
\end{equation}
In our calculations,  we neglect the 1-loop corrections  to the proper
vertices  $L  \tilde{\Phi}   N_\alpha$,  whose  absorptive  parts  are
numerically insignificant in RL. In terms of the absorptive transition
amplitudes  $A^l_{\alpha\beta}$ given in~(\ref{eq:Absl}),  the partial
decay widths $\Gamma_{\alpha l}^{\phantom{C}}$ and their CP-conjugates
$\Gamma_{\alpha l}^C$ may now be expressed in the compact form:
\begin{equation}
\label{Widths}
	\Gamma_{\alpha l}^{\phantom{C}}\ =\  
	m_{N_\alpha} A^l_{\alpha\alpha}\big(\overline{\bf h}^\nu\big)\; , \qquad 
	\Gamma_{\alpha l}^C\ =\ 
	m_{N_\alpha} A^l_{\alpha\alpha}\big(\overline{\bf h}^{\nu C}\big)\; ,
\end{equation}
where we  have explicitly indicated  the dependence of  the absorptive
transition  amplitudes on  $\overline{\bf  h}^\nu$ and  $\overline{\bf
  h}^{\nu C}$.   Given the analytic  expressions~(\ref{Widths}), it is
straightforward   to  compute  the   leptonic  asymmetries   for  each
individual lepton flavour:
\begin{equation}
\label{deltaN}
	\delta_{\alpha l}\ \equiv\ 
	\frac{ 
		\Gamma_{\alpha l}^{\phantom{C}}\: -\:
		\Gamma_{\alpha l}^C 
	}{ 
		\sum\limits_{l = e,\mu ,\tau}
		\Big(\Gamma _{\alpha l}^{\phantom{C}}\: +\:
		\Gamma_{\alpha l}^C\Big)
	}\ =\
	\frac{\big|\overline{\bf h}^\nu_{l\alpha}\big|^2\: -\:
			\big|\overline{\bf h}^{\nu C}_{l\alpha}\big|^2}
	{ 
	\big(\overline{\bf h}^{\nu \dagger}\, 
	\overline{\bf h}^{\nu \phantom{\dagger}}\!\!\big)_{\alpha\alpha}\: +\: 
	\big(\overline{\bf h}^{\nu C\dagger}\, 
	\overline{\bf h}^{\nu C\phantom{\dagger}}\!\!\big)_{\alpha\alpha}
	}\ .
\end{equation}
The analytic results for  the leptonic asymmetries $\delta_{\alpha l}$
  simplify  considerably in  the  2-heavy neutrino  mixing limit,  in
  which $R_{\alpha\beta}$ defined in~(\ref{eq:Rab}) is set to zero. In
  this limit, $\delta_{\alpha l}$ are given by~\cite{APRD,PU}
\begin{equation}
  \label{dCPla}
\delta_{\alpha l} \ \approx\ \frac{{\rm Im} 
\big[\, ({\bf h}^{\nu\dagger}_{\alpha l}{\bf h}^{\nu\phantom{\dagger}}_{l\beta})\,
({\bf h}^{\nu\dagger}_{\phantom{l}}{\bf
    h}^{\nu\phantom{\dagger}}_{\phantom{l}}\!\!)_{\alpha\beta}\big]} 
{({\bf     h}^{\nu\dagger}_{\phantom{l}}
{\bf h}^{\nu\phantom{\dagger}}_{\phantom{l}}\!\!)_{\alpha\alpha}\, ({\bf
    h}^{\nu\dagger}_{\phantom{l}}
{\bf h}^{\nu\phantom{\dagger}}_{\phantom{l}}\!\!)_{\beta\beta}}\
\frac{(m^2_{N_\alpha} - m^2_{N_\beta})\, m_{N_\alpha}\,
  \Gamma^{(0)}_{N_\beta}}{ (m^2_{N_\alpha} - m^2_{N_\beta})^2\: +\: 
m^2_{N_\alpha} \Gamma^{(0)2}_{N_\beta}}\ ,
\end{equation}
where $\alpha, \beta = 1,2$ and $\Gamma^{(0)}_{N_a}$ is the tree-level
decay  width of  the heavy  Majorana neutrino~$N_\alpha$,  given after
(\ref{eq:Abs}). Based on  the simplified expression~(\ref{dCPla}), the
following  two  conditions  for  having resonantly  enhanced  leptonic
asymmetries   $\delta_{\alpha   l}    \sim   {\cal   O}(1)$   may   be
derived~\cite{APRD}:
\begin{eqnarray}
  \label{ResCond}
\mbox{(i)}&& |m_{N_\alpha} -
  m_{N_\beta}|\  \sim\ \frac{\Gamma_{N_{\alpha,\beta}}}{2}\ ,\\
\mbox{(ii)}&& \frac{\big| {\rm Im} 
\big[\, ({\bf h}^{\nu\dagger}_{\alpha l}{\bf h}^{\nu\phantom{\dagger}}_{l\beta})\,
({\bf h}^{\nu\dagger}_{\phantom{l}}{\bf
    h}^{\nu\phantom{\dagger}}_{\phantom{l}}\!\!)_{\alpha\beta}\big]\big|} 
{({\bf     h}^{\nu\dagger}_{\phantom{l}}
{\bf h}^{\nu\phantom{\dagger}}_{\phantom{l}}\!\!)_{\alpha\alpha}\, ({\bf
    h}^{\nu\dagger}_{\phantom{l}}
{\bf
  h}^{\nu\phantom{\dagger}}_{\phantom{l}}\!\!)_{\beta\beta}}\ \sim\ 1\ .
\end{eqnarray}
Note that  the first resonant  condition~(i) is exactly met,  when the
unitarity  limit  on   the  resummed  heavy-neutrino  propagator  gets
saturated~\cite{APNPB},    i.e.~when    the   regulating    expression
in~(\ref{dCPla}),
\begin{equation}
  \label{freg}
f_{\rm reg} \ \equiv\ \frac{| m^2_{N_\alpha} - m^2_{N_\beta} |\: m_{N_\alpha}\,
  \Gamma^{(0)}_{N_\beta}}{ (m^2_{N_\alpha} - m^2_{N_\beta})^2\: +\: 
m^2_{N_\alpha} \Gamma^{(0)2}_{N_\beta}}\ \leq\ 1\,,
\end{equation}
takes its  maximal possible value: $f_{\rm  reg} = 1$.   Within our RL
scenarios,  the   first  condition  in~(\ref{ResCond})   is  naturally
fulfilled as  the heavy-neutrino mass splittings are  generated via RG
effects  and are  of  the  required order.   The  second condition  is
crucial as  well and controls  the size of the  leptonic asymmetries.
As  we will  see below,  the condition~(ii)  in~(\ref{ResCond})  has a
non-trivial impact on approximate $L$-conserving RL models.

\subsection{Comparison with Other Methods}

It  is now  worth  commenting on  some  of the  attempts  made in  the
literature~\cite{FPSW,BP,ABP}  to calculate the  resonant part  of the
leptonic asymmetries~$\delta_{\alpha l}$.  Their results differ by the
way in  which the singularity  $m_{N_2} \to m_{N_1}$ occurring  in the
denominator  of the second  fraction in~(\ref{dCPla})  gets regulated,
when  heavy-neutrino  width  effects  are  taken  into  consideration.
Specifically, the  various approaches differ in  their derivations for
the  analytic  form  of  $f_{\rm  reg}$  given  in~(\ref{freg}).   For
instance,    the   authors    of~\cite{FPSW}   use    a   perturbative
quantum-mechanical approach to obtain a regulator of the form:
\begin{equation}
  \label{Paschos}
f_{\rm reg}\ =\ \frac{\Delta m_N\, \Gamma^{(0)}_{N_{1,2}}/2}{(\Delta
  m_N)^2\: +\:  
\frac{1}{(16\pi)^2}\: m^2_N {\rm Re}^2({\bf h}^{\nu\dagger} {\bf
  h}^\nu)_{12}}\ , 
\end{equation}
where  $m_N =  \frac{1}2 (  m_{N_1}  + m_{N_2}  )$ and  $\Delta m_N  =
|m_{N_1} -  m_{N_2}|$. It is easy  to observe that  for scenarios, for
which  ${\rm Re}({\bf  h}^{\nu\dagger} {\bf  h}^{\nu})_{12} =  0$, but
${\rm           Re}({\bf           h}^{\nu\dagger}_{l1}           {\bf
  h}^{\nu\phantom{\dagger}}_{l2}) \neq  0$, the unitarity  upper bound
given in~(\ref{freg}) gets  violated, in the degenerate heavy-neutrino
mass limit $m_{N_2}  \to m_{N_1}$.  In particular, in  the same limit,
the  individual  lepton-flavour  asymmetries $\delta_{1l}$  (with  $l=
e,\mu,\tau$)  become singular.   Although this  singularity disappears
when the  lepton-flavour sum $\delta_{N_{1,2}}  = \sum_{l=e,\mu ,\tau}
\delta_{1,2l}$ is  taken, the regulator~(\ref{Paschos})  will still be
inapplicable to lepton-flavour RL scenarios, for which $\Delta m_N/m_N
\sim  {\rm  Re}({\bf  h}^{\nu\dagger}  {\bf  h}^\nu)_{12}/(16\pi)  \ll
\Gamma_{N_{1,2}}/m_N$.

Based on a modified version of the field-theoretic approach introduced
in~\cite{APRD,PU},  the authors  of~\cite{BP,ABP}  obtain a  different
regulating expression for the leptonic asymmetry $\delta_{1l}$:
\begin{equation}
  \label{BP}
f_{\rm reg}\ =\ \frac{\big| m^2_{N_1} - m^2_{N_2}\big|\, m_{N_1}\,
  \Gamma^{(0)}_{N_2}}{ \big( m^2_{N_1} - m^2_{N_2}\big)^2\: +\: 
\big( m_{N_1} \Gamma^{(0)}_{N_1} - m_{N_2}\Gamma^{(0)}_{N_2}\big)^2}\ .
\end{equation}
It  is  not  difficult  to  observe that  $f_{\rm  reg}$  diverges  as
$(m_{N_1}-m_{N_2})^{-1}$   in   RL   scenarios,   for   which   $({\bf
  h}^{\nu\dagger}  {\bf h}^{\nu})_{11}  =  ({\bf h}^{\nu\dagger}  {\bf
  h}^{\nu})_{22}$,   even   though  one   could   still  have   $({\bf
  h}^{\nu\dagger}_{l1} {\bf h}^{\nu\phantom{\dagger}}_{l1}) \neq ({\bf
  h}^{\nu\dagger}_{l2} {\bf  h}^{\nu\phantom{\dagger}}_{l2})$ for each
single  lepton  flavour  $l$.   For  instance, such  a  situation  can
naturally occur  in approximate lepton-number conserving  models of RL
discussed recently  in~\cite{BHJ,BBM}.  Evidently, the  prediction for
the leptonic asymmetries $\delta_{N_{1,2}}$  in such scenarios may get
overestimated by many orders of magnitude.

In  order   to  illustrate  this   last  point,  let  us   consider  a
one-generation  inverse   seesaw  model~\cite{MV}  with   two  singlet
neutrinos of  opposite lepton number, $\nu_R$ and  $S_L$. The neutrino
sector  of this  model is  described by  the  lepton-number-conserving
Lagrangian
\begin{equation}
  \label{inverse}
-\, {\cal L}_{\rm inverse}\ =\ 
\frac{1}{2}\, \Big( \bar{S}_L,\ \bar{\nu}^C_R\Big)\,
\left(\!\begin{array}{cc}
0 & M \\ M & 0 \end{array}\!\right)\, 
\left(\!\begin{array}{c} S^C_L \\ \nu_R\end{array}\!\right)\
+\ h_R\,  \Big( \bar{\nu}_L,\ \bar{l}_L\Big)\, \tilde{\Phi} \nu_R\ 
+\ {\rm H.c.}
\end{equation}
Without loss of generality, the kinematic parameters $M$ and $h_R$ can
be  rephased  to  become   real.   Following  closely  the  discussion
in~\cite{APRD}, we  introduce into the  Lagrangian~(\ref{inverse}) the
lepton-number       violating       operators      $\frac{1}2\,\mu_R\,
\bar{\nu}^C_R\nu^{\phantom{C}}_R$,  $\frac{1}2\,  \mu_L\,  \bar{S}^C_L
S^{\phantom{C}}_L$ and  $h_L\, (\bar{\nu}_L,\, \bar{l}_L) \tilde{\Phi}
S^C_L$. In order to minimally break  both the lepton number and CP, it
was  shown in~\cite{APRD}  that  at least  two  of the  aforementioned
$\Delta L =2$ operators are needed.   In fact, this result is a direct
consequence of the  Nanopoulos--Weinberg (NW) no-go theorem~\cite{NW}.
The NW theorem states that no net baryon asymmetry can be generated by
a single $B$- and CP-violating  operator to all orders in perturbation
theory.

It  is interesting  to provide  an  estimate of  the lepton  asymmetry
obtained   within  a   simple  model   of   approximate  lepton-number
conservation, where $\mu_L = -i\mu_R =  \mu$ and $h_L = 0$.  For these
parameters,  we adopt the  same ballpark  of values  as in~\cite{BHJ}:
$m_N  \approx M  = 1$~TeV,  $h_R =  3\times 10^{-2}$  and  $\Delta m_N
\approx  \mu/\sqrt{2}  =  2\times  10^{-10}\: M$.   With  these  input
parameters,  we  may  estimate  that  $\Gamma^{(0)}_{N_{1,2}}  \approx
2\times 10^{-5}  m_N$, leading to  $\Delta m_N /\Gamma^{(0)}_{N_{1,2}}
\approx  10^{-5}$.   The first  fraction  containing the  CP-violating
phases  in~(\ref{dCPla}) is  rather suppressed,  of order  $\mu/M \sim
\Delta m_N/m_N \sim 10^{-10}$, whilst $f_{\rm reg} \approx 2\Delta m_N
/\Gamma_{N_{1,2}}  \approx  2\times  10^{-5}$ within  our  resummation
approach.  This gives rise to dismally small lepton asymmetries:
\begin{equation}
\delta_{1l}\ \approx\ \delta_{2l}\ \sim\ \frac{\mu}{M}\:
\frac{\mu}{\Gamma_{N_{1,2}}}\ \sim \ 10^{-15}\ .
\end{equation}
Notice that the lepton asymmetries $\delta_{1,2l}$ are proportional to
$\mu^2$ in  agreement with  the NW theorem  and, as they  should, both
vanish identically  in the  $L$-conserving limit $\mu  \to 0$.   As we
will see in  the next section, lepton asymmetries  of order $10^{-15}$
will fall short by at least  7 orders of magnitude to explain the BAU.
Instead, had  one used the regulator $f_{\rm  reg}$ in~(\ref{BP}), one
would  have obtained  the enormous  enhancement: $f_{\rm  reg} \approx
\Gamma_{N_{1,2}} /\mu \sim 10^5  \gg 1$, leading to lepton asymmetries
of  order~\cite{BHJ}:  $\delta_{1,2l}  \sim \Gamma_{N_{1,2}}/m_N  \sim
10^{-5}$.  However,  this result is independent  of the $L$-conserving
parameter $\mu = \mu_L = -i\mu_R$, and taking the $L$-conserving limit
$\mu \to  0$, one obtains  a non-zero lepton asymmetry,  which clearly
signifies an erroneous result.  The  above exercise shows that one has
to  go beyond  the two-heavy  neutrino mixing  framework  and consider
non-trivial lepton flavour effects~\cite{APtau},  in order to obtain a
phenomenologically relevant model that predicts testable rates for LFV.

\subsection{Baryon Asymmetry}\label{sec:BaryonAsymmetry}

In  this section  we present  the relevant  Boltzmann  equations (BEs)
which  will  be used  to  evaluate  the heavy-Majorana-neutrino-,  the
lepton-   and  the   baryon-number   densities,  $\eta_{N_{1,2,3}}   =
n_{N_{1,2,3}}/n_\gamma$,  $\eta_{L_{e,\mu,\tau}}  = n_{L_{e,\mu,\tau}}
/n_\gamma$  and~$\eta_B  =  n_B/n_\gamma$,  normalised to  the  photon
number density~$n_\gamma$.   In our computations, we include the
dominant  collision  terms related  to  the  $1\to  2$ decays  of  the
right-handed  neutrinos and  to resonant  $2\to 2$  lepton scatterings
that  describe $\Delta  L  =2$ and  $\Delta  L =  0$ transitions.   We
neglect chemical potential contributions from the right-handed charged
leptons and quarks.

In order to appropriately take into account the flavour effects on the
BEs  for $\eta_{N_{1,2,3}}$  and  $\eta_{L_{e,\mu,\tau}}$, we  closely
follow the approach and  the notation established in~\cite{PU2}. More
explicitly, the BEs are found to be
\begin{eqnarray}
  \label{BEN}
\frac{d\eta_{N_\alpha}}{dz} & = &
	-\ \frac{z}{n_\gamma H_N}\; 
	\bigg( \frac{\eta_{N_\alpha}}{\eta^{\rm eq}_N} - 1 \bigg)\,
	\gamma^{N_\alpha}_{L\Phi}\; , \\[2mm]
  \label{BEDL}
\frac{d\eta_{L_l}}{dz} & = &
	\frac{z}{n_\gamma H_N}\
     \bigg[\; \sum_{\alpha=1}^3
   \bigg(\frac{\eta_{N_\alpha}}{\eta^{\rm eq}_N} - 1\bigg)\, 
        \delta_{\alpha l}\gamma^{N_\alpha}_{L\Phi}\ -\ 
  \frac{2}{3}\; \eta_{L_l}\!\sum_{k=e,\mu,\tau} \bigg(
			\gamma^{L_l\Phi}_{L^C_k\Phi^\dagger} +
			\gamma^{L_l\Phi}_{L_k\Phi} \bigg)\nonumber\\
&& -\ \frac{2}{3}\sum_{k=e,\mu,\tau}\, \eta_{L_k}
 \bigg(	\gamma^{L_k\Phi}_{L^C_l\Phi^\dagger} -
 \gamma^{L_k\Phi}_{L_l\Phi} \bigg)\; \bigg]\; ,
\end{eqnarray}
where  $\alpha = 1,\,2,\,3$  and $l  = e,\,\mu,\,\tau$.   In addition,
$H_N\approx  17\times  m_N^2/M_{\rm P}$  is  the  Hubble parameter  at
$T=m_N$, where $M_{\rm P} = 1.2\times 10^{19}$~GeV is the Planck mass.
The   $T$-dependence  of   the  BEs~(\ref{BEN})   and~(\ref{BEDL})  is
expressed by virtue of the dimensionless parameter
\begin{equation}
  \label{zparam}
	z\ =\ \frac{m_N}{T}\ ,
\end{equation}
in terms of which the photon number density is given by
\begin{equation}
	n_\gamma\ =\ \frac{2T^3}{\pi^2}\; \zeta(3)\ =\ 
	\frac{2m_N^3}{\pi^2}\; \frac{\zeta(3)}{z^3}\ ,
\end{equation}
where  $\zeta(3)\approx   1.202$  is  Ap\'ery's   constant.   Finally,
$\eta_N^{\rm eq}$  in~(\ref{BEN}) and (\ref{BEDL})  is the equilibrium
number density  of the heavy Majorana  neutrino $N_\alpha$, normalised
to the number density of photons, i.e.
\begin{equation}
\label{etaNeq}
	\eta_N^{\rm eq}\ \approx\ \frac{1}{2}z^2 
	K_2\left(z\right),
\end{equation}
where    $K_n    (x)$    is    the   $n$th-order    modified    Bessel
function~\cite{AS}. 

The BEs~(\ref{BEN})  and (\ref{BEDL}) include the  collision terms for
the  decays $N_\alpha  \to L_l\Phi$,  as  well as  the $\Delta  L=0,2$
resonant scatterings:  $L_k\Phi \leftrightarrow L_l\Phi$  and $L_k\Phi
\leftrightarrow L^C_l\Phi^\dagger$, which are defined as~\cite{PU2}
\begin{eqnarray}
  \label{gammaN}
\gamma^{N_\alpha}_{L\Phi} & \equiv &
		\sum_{k = e,\mu,\tau} \Big[
		  \gamma(N_\alpha \to L_k\Phi) 
		+ \gamma(N_\alpha \to L^C_k\Phi^\dagger)\,\Big]\;,\nonumber\\
  \label{DL0}
\gamma^{L_k\Phi}_{L_l\Phi} & \equiv &  	\gamma(L_k\Phi \to L_l\Phi)\:
+\: \gamma(L^C_k\Phi^\dagger \to L^C_l\Phi^\dagger)\; ,\\[2mm] 
  \label{DL2}
\gamma^{L_k\Phi}_{L^C_l\Phi^\dagger} & \equiv &  
   \gamma(L_k\Phi \to L^C_l\Phi^\dagger)\: +\: 
                         \gamma(L^C_k\Phi^\dagger \to L_l\Phi )\; .\nonumber
\end{eqnarray}
Since we are only interested in  the resonant part of the above $2 \to
2$ scatterings,  we make use  of the narrow width  approximation (NWA)
for the  resummed heavy-neutrino propagators.  Thus,  at the amplitude
squared level, we employ the  NWA for the complex-conjugate product of
Breit--Wigner propagators~\cite{Frank}:
\begin{align}
  \label{NWA}
	\frac{1}{s-m_{N_\alpha}^2-i m_{N_\alpha}\Gamma_{N_\alpha}}
	\frac{1}{s-m_{N_\beta}^2+i m_{N_\beta}\Gamma_{N_\beta}}\
	\approx\
	\frac{i\pi\left[\delta(s-m_{N_\alpha}^2) + \delta(s-m_{N_\beta}^2)\right]}{
		m_{N_\alpha}^2-m_{N_\beta}^2 + \frac{i}{2}
		(m_{N_\alpha}+m_{N_\beta})(\Gamma_{N_\alpha}+\Gamma_{N_\beta})}\ .
\end{align}
Note  that the  NWA relation~(\ref{NWA})  is  valid for  any range  of
parameter values for $\Gamma_{N_{1,2,3}}$ and $m_{N_{1,2,3}}$, as long
as  $\Gamma_{N_{1,2,3}} \ll  m_{N_{1,2,3}}$.  This  last  condition is
naturally fulfilled within the R$\ell$L models.

The  collision term  pertinent to  the  heavy-Majorana-neutrino decays
$\gamma^{N_\alpha}_{L\Phi}$ is given by
\begin{equation}
  \label{eq:gamma1}
\gamma^{N_\alpha}_{L\Phi}\	=\ \frac{m_N^3}{\pi^2
  z}\; K_1(z)\, \Gamma_{N_\alpha}\; ,
\end{equation}
whilst  the corresponding  collision terms  for the  $\Delta L  = 0,2$
lepton-flavour   transitions   are   calculated   by  means   of   the
NWA~(\ref{NWA}) to be
\begin{align}
  \label{eq:gamma2}
	\gamma^{L_k\Phi}_{L_l\Phi}\ &=\ \sum_{\alpha,\beta=1}^3
		\left(\gamma^{N_\alpha}_{L\Phi}+\gamma^{N_\beta}_{L\Phi} \right) 
		\frac{2\left(
			\overline{\bf h}^{\nu*}_{l\alpha}\,
			\overline{\bf h}^{\nu C*}_{k\alpha}\,
			\overline{\bf h}^{\nu}_{l\beta}\,
			\overline{\bf h}^{\nu C}_{k\beta}\:
			+\:
			\overline{\bf h}^{\nu C*}_{l\alpha}\,
			\overline{\bf h}^{\nu*}_{k\alpha}\,
			\overline{\bf h}^{\nu C}_{l\beta}\,
			\overline{\bf h}^{\nu}_{k\beta}
		\right)}{\left[
			(\overline{\bf h}^{\nu \dagger}
			\overline{\bf h}^{\nu})_{\alpha\alpha}\: +\:
			(\overline{\bf h}^{\nu C\dagger}
			\overline{\bf h}^{\nu C})_{\alpha\alpha}\: +\:
			(\overline{\bf h}^{\nu \dagger}
			\overline{\bf h}^{\nu})_{\beta\beta}\: +\:
			(\overline{\bf h}^{\nu C\dagger}
			\overline{\bf h}^{\nu C})_{\beta\beta}
		\right]^2} \nonumber\\
		&\qquad\,\times 
		\left(
         1\: -\: 2i\frac{m_{N_\alpha}-m_{N_\beta}}{\Gamma_{N_\alpha}+\Gamma_{N_\beta}} 
		\right)^{-1}\; , \\
\label{eq:gamma3}
	\gamma^{L_k\Phi}_{L_l^C\Phi^\dagger}\ &=\ \sum_{\alpha,\beta=1}^3
		\left(\gamma^{N_\alpha}_{L\Phi}+\gamma^{N_\beta}_{L\Phi} \right) 
		\frac{2\left(
			\overline{\bf h}^{\nu*}_{l\alpha}\,
			\overline{\bf h}^{\nu*}_{k\alpha}\,
			\overline{\bf h}^{\nu}_{l\beta}\,
			\overline{\bf h}^{\nu}_{k\beta}\:
			+\:
			\overline{\bf h}^{\nu C*}_{l\alpha}\,
			\overline{\bf h}^{\nu C*}_{k\alpha}\,
			\overline{\bf h}^{\nu C}_{l\beta}\,
			\overline{\bf h}^{\nu C}_{k\beta}
		\right)}{\left[
			(\overline{\bf h}^{\nu \dagger}
			\overline{\bf h}^{\nu})_{\alpha\alpha}\: +\:
			(\overline{\bf h}^{\nu C\dagger}
			\overline{\bf h}^{\nu C})_{\alpha\alpha}\: +\:
			(\overline{\bf h}^{\nu \dagger}
			\overline{\bf h}^{\nu})_{\beta\beta}\: +\:
			(\overline{\bf h}^{\nu C\dagger}
			\overline{\bf h}^{\nu C})_{\beta\beta}
		\right]^2} \nonumber\\
		&\qquad\,\times 
		\left(
    1\: -\: 2i\frac{m_{N_\alpha}-m_{N_\beta}}{\Gamma_{N_\alpha}+\Gamma_{N_\beta}}
		\right)^{-1}\; .
\end{align}
The scattering collision terms (\ref{eq:gamma2}) and (\ref{eq:gamma3})
contain all  contributions from the resonant  exchange of right-handed
neutrinos in the NWA,  including contributions from the so-called real
intermediate  states (RISs)~\cite{KW}.  The  latter are  obtained when
only the  diagonal terms $\alpha =  \beta$ are taken  in the summation
over the heavy Majorana states $N_{\alpha,\beta}$.

\vfill\eject

Separating the  diagonal $\alpha =  \beta$ RIS contributions  from the
off-diagonal $\alpha \neq \beta$ terms  in the sum, we may rewrite the
BE~(\ref{BEDL}) in the form~\cite{PU2}
\begin{eqnarray}
  \label{BEDL4}
	\frac{d\eta_{L_l}}{dz} \!\!& = &\!\!
	\frac{z}{n_\gamma H_N}
	\bigg\{\;
		\sum_{\alpha=1}^3
		\bigg(
			\frac{\eta_{N_\alpha}}{\eta^{\rm eq}_N} - 1
		\bigg)\, \delta_{\alpha l}\gamma^{N_\alpha}_{L\Phi}\:
		-\: \frac{2}{3}\,\eta_{L_l}\bigg[\,
		\sum_{\alpha=1}^3 \gamma^{N_\alpha}_{L\Phi} B_{\alpha l} + 
		\sum_{k=e,\mu,\tau}
		\bigg( {\gamma'}^{L_l\Phi}_{L^C_k\Phi^\dagger} +
			{\gamma'}^{L_l\Phi}_{L_k\Phi}
		\bigg)\,\bigg] \nonumber\\
	&&
	-\ \frac{2}{3}\!\sum_{k=e,\mu,\tau}\!\!\eta_{L_k}
		\bigg[\,
		\sum_{\alpha=1}^3
		\delta_{\alpha l}\delta_{\alpha k} \gamma^{N_\alpha}_{L\Phi}\: +\:
		\bigg(	{\gamma'}^{L_k\Phi}_{L^C_l\Phi^\dagger} -
			{\gamma'}^{L_k\Phi}_{L_l\Phi}
		\bigg)\, \bigg]\, \bigg\}\; ,
\end{eqnarray}
where
${\gamma'}^{X}_{Y}={\gamma}^{X}_{Y}-({\gamma}^{X}_{Y})_\text{RIS}$
denote the  RIS-subtracted collision terms and $B_{\alpha  l}$ are the
branching ratios
\begin{equation}
	B_{\alpha l}\ = \
	\frac{ 
		\Gamma_{\alpha l}^{\phantom{C}}\: +\:
		\Gamma_{\alpha l}^C 
	}{ 
		\sum\limits_{k = e,\mu ,\tau}
		\left(\Gamma _{\alpha k}^{\phantom{C}}\: +\:
		\Gamma_{\alpha k}^C\right)
	}\ =\
	\frac{\big|\overline{\bf h}^\nu_{l\alpha}\big|^2\: +\:
			\big|\overline{\bf h}^{\nu C}_{l\alpha}\big|^2}
	{ 
	\big(\overline{\bf h}^{\nu \dagger}\, 
	\overline{\bf h}^{\nu \phantom{\dagger}}\!\!\big)_{\alpha\alpha}\: +\: 
	\big(\overline{\bf h}^{\nu C\dagger}\, 
	\overline{\bf h}^{\nu C\phantom{\dagger}}\!\!\big)_{\alpha\alpha}
	}\ .
\end{equation}
The   collision   terms    proportional   to   $\eta_{L_k}$   and   to
$\delta_{\alpha l}\delta_{\alpha k}$  on the RHS of~(\ref{BEDL4}) turn
out  to be  numerically negligible  for the  R$\ell$L  scenarios under
consideration.    Instead,    the   RIS-subtracted   collision   terms
proportional  to  $\eta_{L_l}$  in~(\ref{BEDL4})  become  significant.
Their importance  in RL models was  originally raised in~\cite{PU,PU2}
and confirmed most recently in~\cite{BHJ}.

The  next step  is  to  include the  effect  of the  $(B+L)$-violating
sphaleron processes~\cite{KRS}.  For  temperatures $T$ larger than the
critical temperature  $T_c \approx  135$~GeV of the  electroweak phase
transition,  the  conversion   of  the  total  lepton-to-photon  ratio
$\sum_{l=e,\mu,\tau}   \eta_{L_l}$  to   the   baryon-to-photon  ratio
$\eta_B^c$ at $T_c$ is given by the relation:
\begin{equation}
  \label{etaBc}
	\eta_B^c\ =\ -\;\frac{28}{51}\; \sum_{l=e,\mu,\tau} \eta_{L_l}\ .
\end{equation}
For  $T <  T_c$,  the so-generated  baryon  asymmetry $\eta_B^c$  gets
diluted by standard photon interactions until the recombination epoch,
leading to the BAU
\begin{equation}
  \label{etaB0}
	\eta_B\ \approx\ \frac{1}{27}\; \eta_B^c.
\end{equation}
This  theoretical prediction  can  now be  compared  with the  current
observational        value       for        the       baryon-to-photon
asymmetry\cite{Komatsu:2010fb},
\begin{equation}
  \label{etaBobs}
   \eta_B^{\rm obs}\ =\ \left(6.20\pm 0.15\right) \times 10^{-10}\; .
\end{equation}
\begin{figure}[t]
\centering
\includegraphics[clip,width=0.90\textwidth]{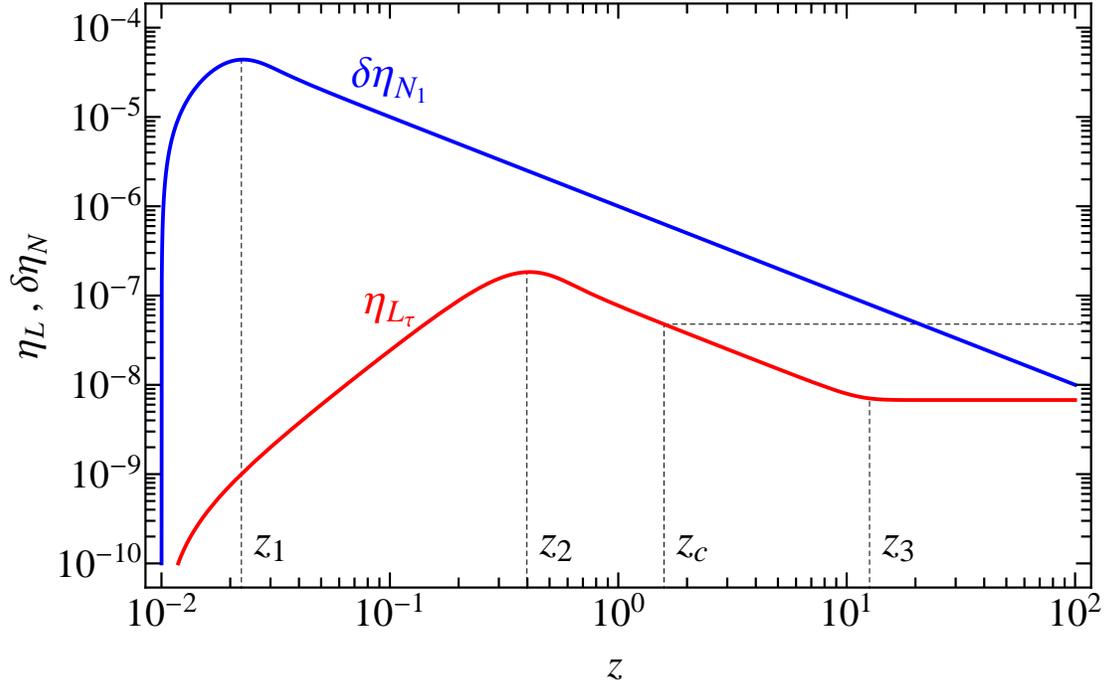}
\caption{\it Numerical solutions to BEs (\ref{BEN2}) and (\ref{BEDL3})
  for  $\delta\eta_{N_1} =  \eta_{N_1}/\eta_N^{\rm eq}  -  1$ and
  $\eta_{L_\tau}$,  respectively, in  a  R$\tau$L model  with $m_N  =
  220$~GeV,  $K_1 =  10^6$,  $K^{\rm eff}_\tau=10^2$, $\delta_\tau=10^{-6}$  and
  $z_c=m_N/T_c=1.6$.}
\label{fig:eta_z} 
\end{figure}
It is instructive  to derive approximate numerical solutions
to the BEs~(\ref{BEN}) and~(\ref{BEDL4}).   To this end, we re-express
the  BEs  in  terms  of the  out-of-equilibrium  deviation  parameters
$\delta\eta_{N_\alpha}  =  \eta_{N_\alpha}/\eta_N^{\rm  eq} -  1$  and
neglect the suppressed $\mathcal{O}(\delta^2_{\alpha l})$ collision
terms.   For  simplicity,  we  initially  ignore  the  RIS  subtracted
collision  terms.  But, as  we  will  see  below, their  inclusion  is
straightforward.  With  these approximations and  simplifications, the
BEs may be recast into the compact form:
\begin{align}
\label{BEN2}
	\frac{d\delta\eta_{N_\alpha}}{dz}\ &=\
	\frac{K_1(z)}{K_2(z)}\;
	\bigg[\, 1\: +\: \Big(1 - K_\alpha z \Big)\,
                                  \delta\eta_{N_\alpha}\bigg]\;, \\
\label{BEDL2}
	\frac{d\eta_{L_l}}{dz}\ &=\
	z^3 K_1(z)\;
	\sum_{\alpha=1}^3
	K_\alpha
	\bigg(
	\delta\eta_{N_\alpha} \delta_{\alpha l}\:
	-\: \frac{2}{3} B_{\alpha l} \eta_{L_l}
	\bigg)\; ,
\end{align}
with $K_\alpha = \Gamma_{N_\alpha}/[\zeta(3)\, H_N]$. In the kinematic
regime  $z  >   z_1^\alpha  \approx  2K_\alpha^{-1/3}$,  the  solution
to~(\ref{BEN2}) may well be approximated by
\begin{equation}
	\delta\eta_{N_\alpha}(z)\ =\ (K_\alpha z)^{-1}\; ,
\end{equation}
independently of the initial conditions (see~Figure~\ref{fig:eta_z}). In
this regime, the BE~(\ref{BEDL2}) becomes
\begin{equation}
\label{BEDL3}
	\frac{d\eta_{L_l}}{dz}\ =\
	z^2 K_1(z)\,
\bigg(\,\delta_l\: -\: \frac{2}{3}\, z K_l\, \eta_{L_l}\,\bigg)\; ,
\end{equation}
with       $\delta_l=\sum_{\alpha=1}^3\delta_{\alpha      l}$      and
$K_l=\sum_{\alpha=1}^3K_\alpha  B_{\alpha  l}$.   We may  include  the
numerically significant RIS-subtracted collision terms proportional to
$\eta_{L_l}$  in~(\ref{BEDL4}),  by rescaling  $K_l  \to \kappa_l  K_l
\equiv K^{\rm eff}_l$, where 
\begin{eqnarray}
  \label{kappal}
\kappa_l \!\!& \equiv &\!\!
\frac{\sum\limits_{k=e,\mu,\tau}
		\bigg( \gamma^{L_l\Phi}_{L^C_k\Phi^\dagger} +
			\gamma^{L_l\Phi}_{L_k\Phi}
		\bigg)\ +\ \gamma^{L_l\Phi}_{L^C_l\Phi^\dagger} -
			\gamma^{L_l\Phi}_{L_l\Phi} }
{\sum\limits_{\alpha=1}^3 \gamma^{N_\alpha}_{L\Phi}\, B_{\alpha l}}\nonumber\\
\!\!& = &\!\!
2 \sum_{\alpha,\beta=1}^3
  \frac{\left(\overline{\bf h}^{\nu\dagger}_{\alpha l}\,
			\overline{\bf h}^{\nu}_{l\beta}\:
			+\:
			\overline{\bf h}^{\nu C\dagger}_{\alpha l}\,
			\overline{\bf h}^{\nu C}_{l\beta}\, \right)\, 
                   \left[ (\overline{\bf h}^{\nu \dagger}
			\overline{\bf h}^{\nu})_{\alpha\beta}\: +\:
			(\overline{\bf h}^{\nu C\dagger}
			\overline{\bf h}^{\nu C})_{\alpha\beta}\right]\
+\
 \left(\overline{\bf h}^{\nu\dagger}_{\alpha l}\,
			\overline{\bf h}^{\nu}_{l\beta}\:
			-\:
			\overline{\bf h}^{\nu C\dagger}_{\alpha l}\,
			\overline{\bf h}^{\nu C}_{l\beta}\, \right)^2}
                        {\left[
            (\overline{\bf h}^{\nu}\overline{\bf h}^{\nu\dagger})_{ll}
                \: +\:
	    (\overline{\bf h}^{\nu C} \overline{\bf h}^{\nu C\dagger})_{ll}\,
		\right]\,
                        \left[
			(\overline{\bf h}^{\nu \dagger}
			\overline{\bf h}^{\nu})_{\alpha\alpha}\: +\:
			(\overline{\bf h}^{\nu C\dagger}
			\overline{\bf h}^{\nu C})_{\alpha\alpha}\: +\:
			(\overline{\bf h}^{\nu \dagger}
			\overline{\bf h}^{\nu})_{\beta\beta}\: +\:
			(\overline{\bf h}^{\nu C\dagger}
			\overline{\bf h}^{\nu C})_{\beta\beta}
		\right]} \nonumber\\
		&&\times 
		\left(
    1\: -\: 2i\frac{m_{N_\alpha}-m_{N_\beta}}{\Gamma_{N_\alpha}+\Gamma_{N_\beta}}
		\right)^{-1}\; .
\end{eqnarray}
In  determining the scaling  factor $\kappa_l$,  we have  assumed that
$\eta_{L_l} \gg \eta_{L_{k\neq l}}$  in~(\ref{BEDL}), which is a valid
approximation within a given R$\ell$L scenario under study.  Note that
if  only the  diagonal $\alpha  =  \beta$ terms  representing the  RIS
contributions  are  considered  in  the sum,  $\kappa_l$  reaches  its
maximum value,  i.e.~$\kappa_l = 1  + {\cal O}(\delta^2_l )$.  We also
have  checked that  in the  $L_l$-conserving  limit of  the theory,  the
parameter $\kappa_l$ vanishes, as it should.

As is  illustrated in Figure~\ref{fig:eta_z},  the solution~$\eta_{L_l}$
to~(\ref{BEDL3}) exhibits  different behaviour in  the three kinematic
regimes, characterized  by the specific  values of the parameter  $z =
m_N/T$:
\begin{equation}
\label{eq:z23}
	z_2^l\ \approx\ 2(K^{\rm eff}_l)^{-1/3}\;,\qquad
	z_3^l\ \approx\ 1.25\, \ln\left(25 K^{\rm eff}_l\right)\ .
\end{equation}
For  $z$  values in  the  range  $z_2^l <  z  <  z_3^l$, the  solution
$\eta_{L_l}$ may well be approximated by
\begin{equation}
  \label{etaL3}
	\eta_{L_l}(z)\ =\ \frac{3}{2}\; \frac{\delta_l}{K^{\rm eff}_l z}\ .
\end{equation}
For $z> z_3^l$, the lepton number density $\eta_{L_l}$ freezes out and
approaches  the  constant  value  $\eta_{L_l}  =  (3\delta_l)/(2K^{\rm
  eff}_l z_3^l)$\footnote{The  onset of  the freeze-out is  defined as
  the position $z_3^l$ where the relative slope of (\ref{BEDL3}) drops
  below~1,     i.e.~when    $|\eta_{L_l}'/\eta_{L_l}|=2/3    (z_3^l)^3
  K_1(z_3^l) K^{\rm  eff}_l=1$.  The solution to this  equation can be
  analytically  expressed in  terms of  the Lambert  $W$  function, to
  which  $z_3^l$   in  (\ref{eq:z23})   proves  to  be   an  excellent
  interpolating approximation  over the  wide range of  values $K^{\rm
    eff}_l\approx   2  -   10^{10}$.}.   The   general   behaviour  of
$\eta_{L_l}$    in   the   different    regimes   is    displayed   in
Figure~\ref{fig:eta_z}.

In  this paper  we only  consider  R$\ell$L scenarios,  for which  the
washout  is   strong  enough,  such  that   the  critical  temperature
$z_c=m_N/T_c$ where  the baryon asymmetry $\eta_B$  decouples from the
lepton asymmetries $\eta_{L_l}$ is  situated in the linear drop-off or
constant regime.  Specifically, we require that
\begin{equation}
  \label{Kzc}
	z_c\ >\ 2\,K_\alpha^{-1/3}\; , \quad
	z_c\ >\ 2\,(K_l^\text{eff})^{-1/3}\; ,
\end{equation}
for  all heavy  neutrino  species $N_\alpha  =  N_{1,2,3}$ and  lepton
flavours  $l=e,\mu,\tau$.   As  a  consequence, the  baryon  asymmetry
$\eta_B$  becomes   relatively  independent  of   the  initial values
of~$\eta_{L_l}$  and~$\eta_{N_\alpha}$.   In  this case,  taking  into
account all factors in~(\ref{etaBc}), (\ref{etaB0}) and (\ref{etaL3}),
the resulting BAU is estimated to be
\begin{eqnarray}
  \label{etaBth}
\eta_B \!&=&\! -\; \frac{28}{51}\;\frac{1}{27}\; \frac{3}{2}\,
  \sum_{l=e,\mu,\tau} \frac{\delta_l}{K^{\rm eff}_l\: {\rm
      min}(z_c,z_3^l)}\nonumber \\
\!&\approx&\!
	-\; 3\cdot 10^{-2}
	\sum_{l=e,\mu,\tau}\,
\frac{\delta_l}{K^{\rm eff}_l\: 
{\rm min}\Big[m_N/T_c\, ,\, 1.25\ln(25K^{\rm eff}_l)\Big]}\; .
\end{eqnarray}
We  note  that  the  formula~(\ref{etaBth})  provides  a  fairly  good
estimate  of the BAU  $\eta_B$ to  less than  20\%, in  the applicable
regime  of  approximations  given  by~(\ref{Kzc}) for  $K^{\rm  eff}_l
\stackrel{>}{{}_\sim} 5$ for a right-handed neutrino mass scale of the
order  of  the EW  scale.   Hence, to  account  for  the observed  BAU
$\eta^{\rm   obs}_B$  given  in~(\ref{etaBobs}),   lepton  asymmetries
$\delta_l  \stackrel{>}{{}_\sim} 10^{-7}$ are  required.  In  the next
section,  we  present numerical  estimates  of  the  BAU for  R$\ell$L
scenarios, based on the simplified BEs~(\ref{BEN2}) and (\ref{BEDL3}),
with  $K_l$   replaced  by  $K^{\rm   eff}_l$  as  defined   by  means
of~(\ref{kappal}).

\setcounter{equation}{0}
\section{Numerical Results}\label{sec:num}

In this section we present numerical estimates of the BAU $\eta_B$ and
the low-energy LFV observables for  $\mu \to e$ and $\tau \to (e,\mu)$
transitions,    based   on   the    analytic   results    derived   in
Sections~\ref{sec:LowEnergyObservables}   and  \ref{sec:Leptogenesis}.
Our aim  is to  delineate the viable  parameter space of  the R$\ell$L
models: R$\tau$L,  R$\mu$L and R$e$L,  by considering both cases  of a
normal and  inverted hierarchy for  the light neutrino  mass spectrum.
In all  the R$\ell$L scenarios  under study, the lightest  neutrino is
massless. As described in Section~\ref{sec:Model}, we use the neutrino
oscillation data to determine  the theoretical parameters of the light
neutrino  mass  matrix~${\bf  m}^\nu$.   Specifically, we  invert  the
seesaw formula  and solve for  the neutrino Yukawa couplings  $a$, $b$
and   $\epsilon_{e,\mu,\tau}$.   In~addition,   the  electroweak-scale
flavour structure of the right-handed neutrino mass matrix~${\bf m}_M$
is generated from a flavour-universal heavy-neutrino mass matrix ${\bf
  m}_M(M_X) = m_N {\bf 1}_3$ at the GUT scale $M_X$, after taking into
account RG-running effects.

\begin{figure}[t]
\centering
\includegraphics[clip,width=0.49\textwidth]{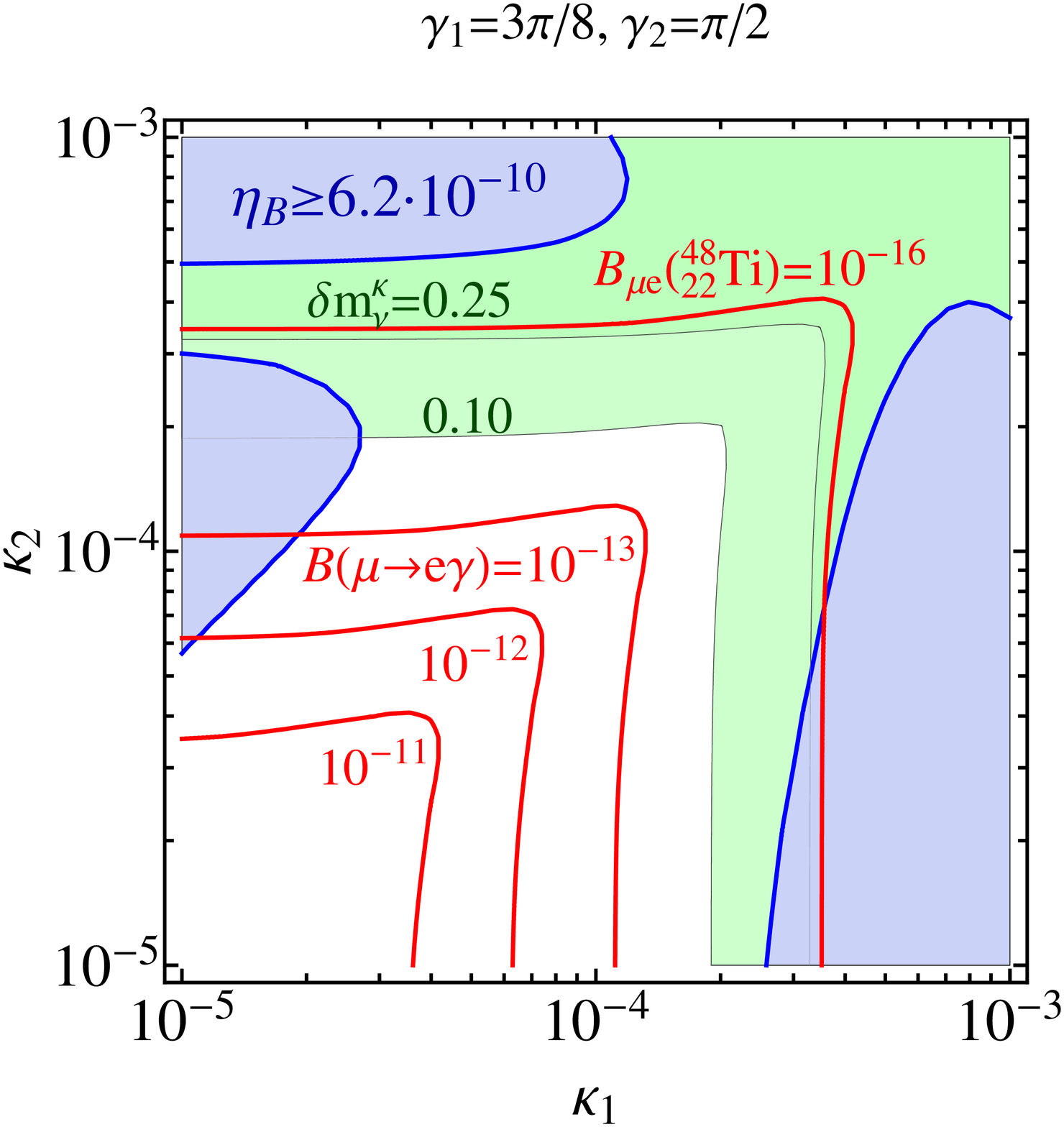}
\includegraphics[clip,width=0.49\textwidth]{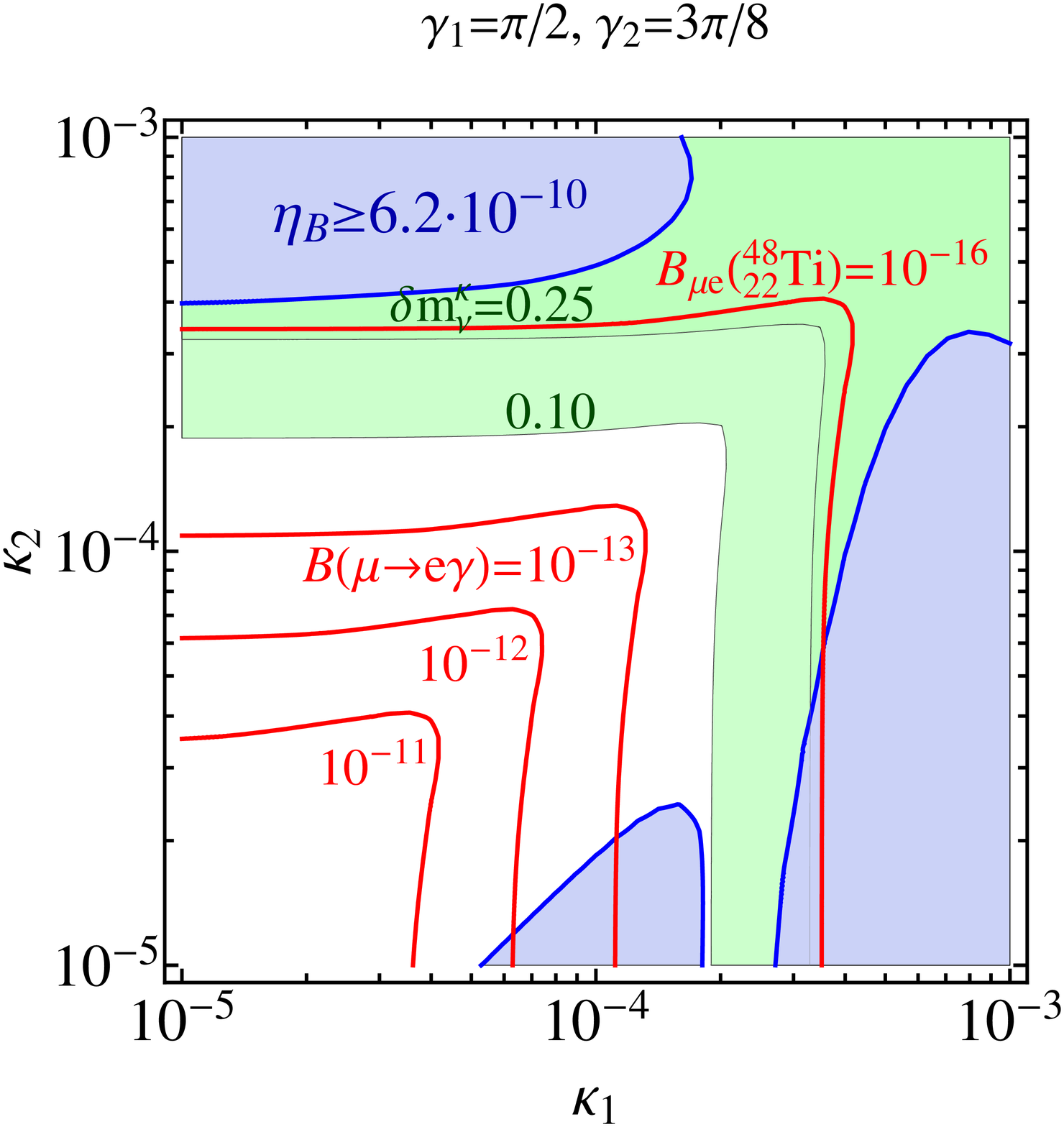}
\caption{\it The baryon asymmetry $\eta_B$ (blue contours) and the LFV
  observables  $B(\mu \to  e\gamma)$  and $B_{\mu  e}({}^{48}_{22}{\rm
  Ti})$ (red  contours) as functions  of $\kappa_1$ and  $\kappa_2$ in
  the  R$\tau$L  model with  $m_N=120$~GeV,  assuming  a normal  light
  neutrino mass  spectrum. The remaining parameters are  chosen to be:
  $\gamma_1=3\pi/8$,   $\gamma_2=\pi/2$,   $\phi_1=\pi$,   $\phi_2=0$,
  $\text{Re}(a)>0$~(left  panel); $\gamma_1=\pi/2$, $\gamma_2=3\pi/8$,
  $\phi_1=0$,   $\phi_2=0$,   $\text{Re}(a)<0$~(right   panel).    The
  neutrino oscillation  parameters are set  at their best  fit values,
  with  $\sin^2\theta_{13}=0.033$ at  its $2\sigma$  upper  limit. The
  blue  shaded regions  denote the  parameter space  where  the baryon
  asymmetry  is   larger  than  the   observational  value  $\eta^{\rm
  obs}_B=6.2\times  10^{-10}$.   The green  shaded  areas labelled  as
  `$\,\delta  {\rm m}_\nu^\kappa=0.25$'  and  `$\,0.10$' indicate  the
  parameter  space  where the  inversion  of  the light-neutrino  mass
  matrix is violated at the 25\% and 10\% level, respectively.}
\label{fig:k1_k2_RtauL_75_100} 
\end{figure}

\begin{figure}[t]
\centering
\includegraphics[clip,width=0.49\textwidth]{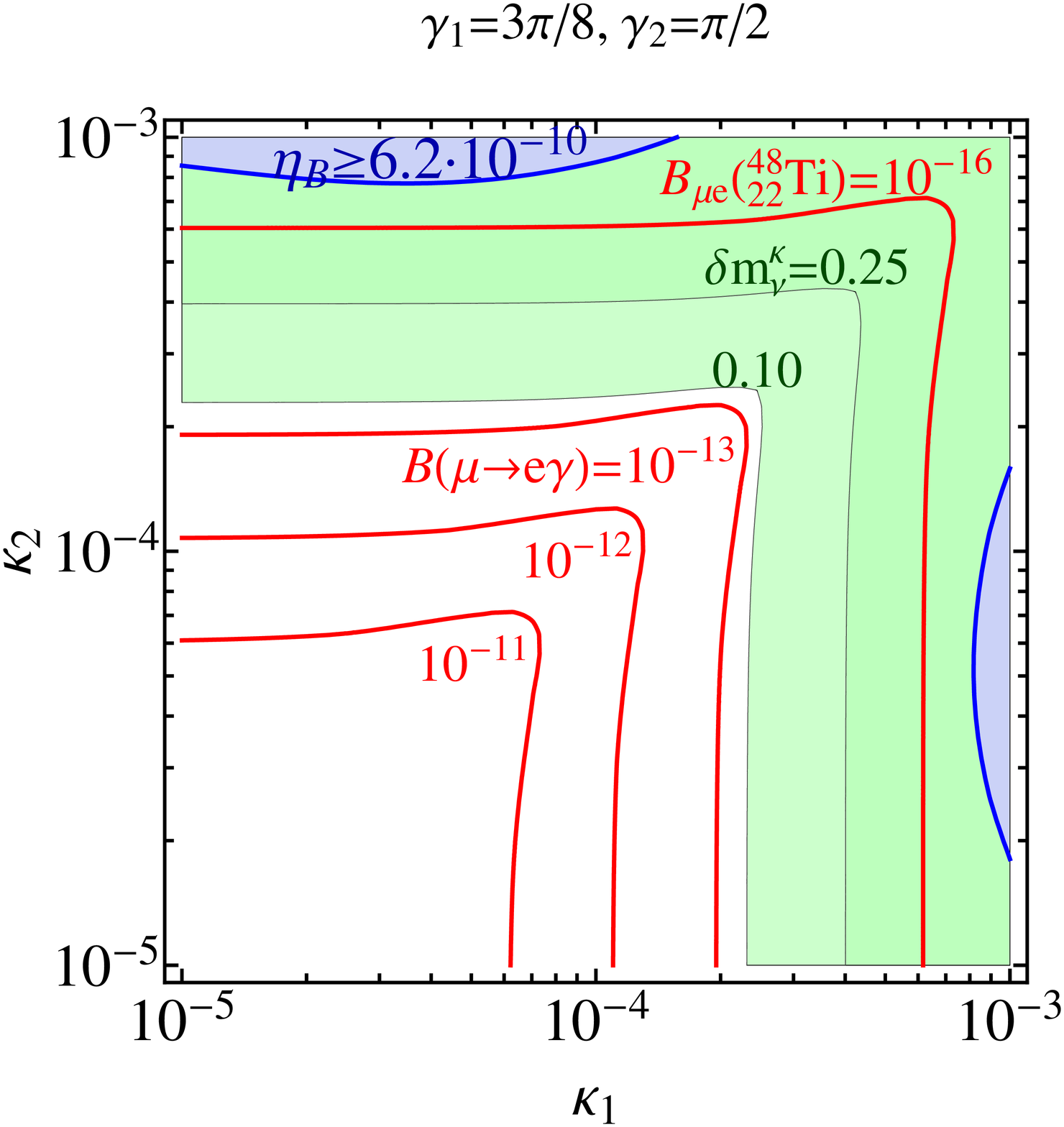}
\includegraphics[clip,width=0.49\textwidth]{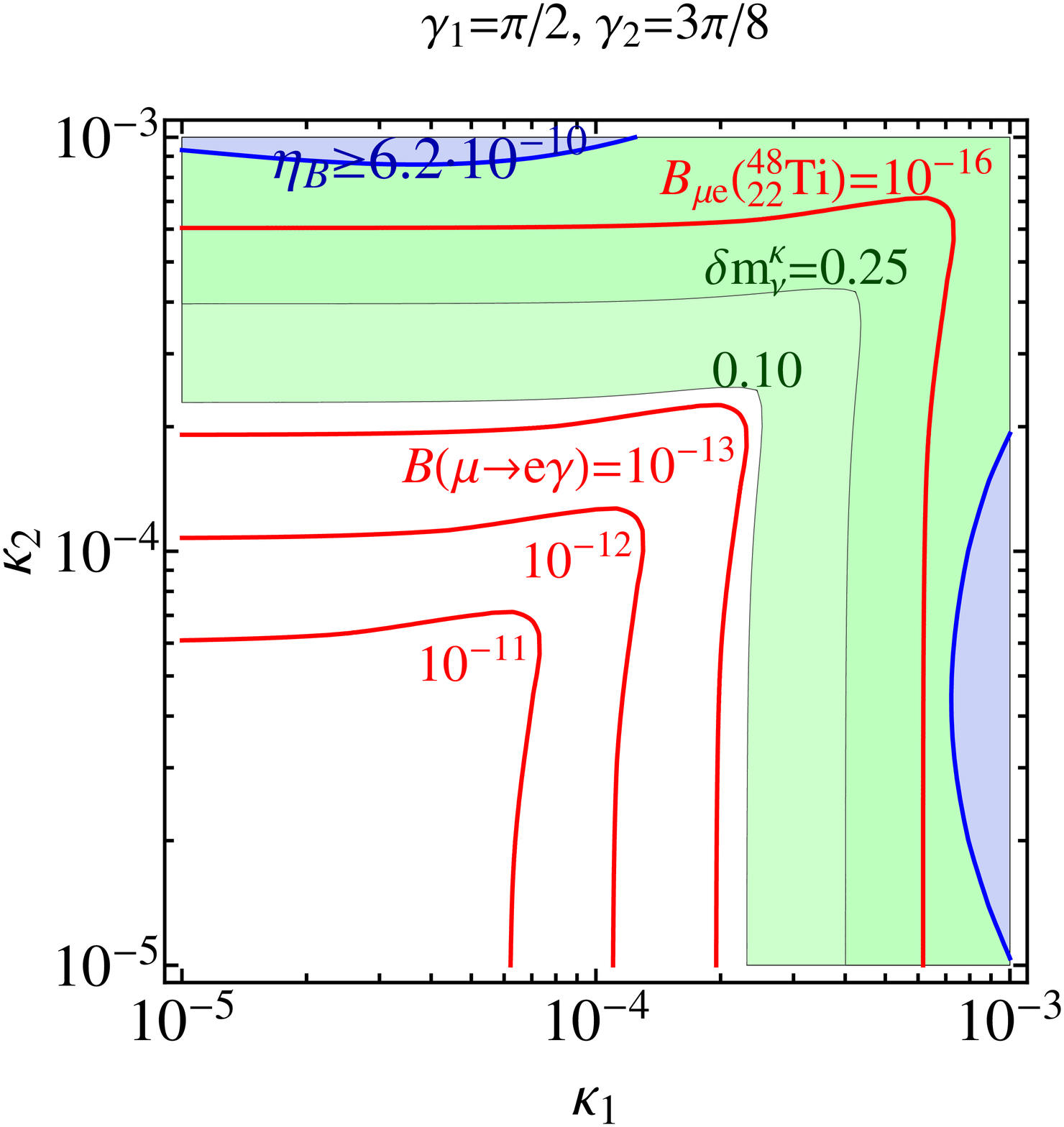}
\caption{\it The same as in Figure~\ref{fig:k1_k2_RtauL_75_100}, but
  assuming an inverted light neutrino mass spectrum. The remaining
  parameters are chosen as follows: $\gamma_1=3\pi/8$,
  $\gamma_2=\pi/2$, $\phi_1=0$, $\phi_2=0$, $\text{Re}(a)>0$~(left
  panel); $\gamma_1=\pi/2$, $\gamma_2=3\pi/8$, $\phi_1=0$, $\phi_2=0$,
  $\text{Re}(a)>0$~(right panel).}
\label{fig:k1_k2_RtauL_75_100_inv} 
\end{figure}

\begin{figure}[t]
\centering
\includegraphics[clip,width=0.8\textwidth]{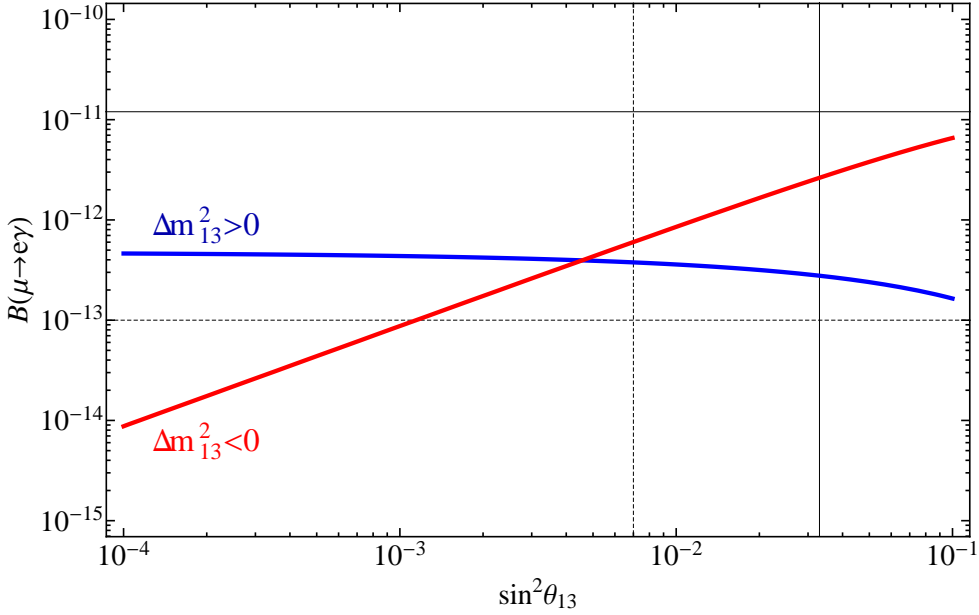}
\caption{\it The branching ratio $B(\mu\to e\gamma)$ as a function of
  $\sin^2\theta_{13}$, in R$\tau$L scenarios with a normal (blue) and
  inverted (red) light neutrino mass spectra. The model parameters for these
  two cases are as in Figures~\ref{fig:k1_k2_RtauL_75_100}~(left panel) and
  \ref{fig:k1_k2_RtauL_75_100_inv}~(left panel), respectively, with
  $\kappa_1=10^{-5}$ and $\kappa_2=10^{-4}$. The horizontal solid
  (dashed) line denotes the current (expected future) limit on
  $B(\mu\to e\gamma)$. The vertical solid and dashed lines denote the
  $2\sigma$ upper limit and the nominal best fit value of
  $\sin^2\theta_{13}$, the latter roughly corresponding to the
  expected sensitivity of future oscillation experiments.}  
\label{fig:theta13} 
\end{figure}

\begin{figure}[t]
\centering
\includegraphics[clip,width=0.8\textwidth]{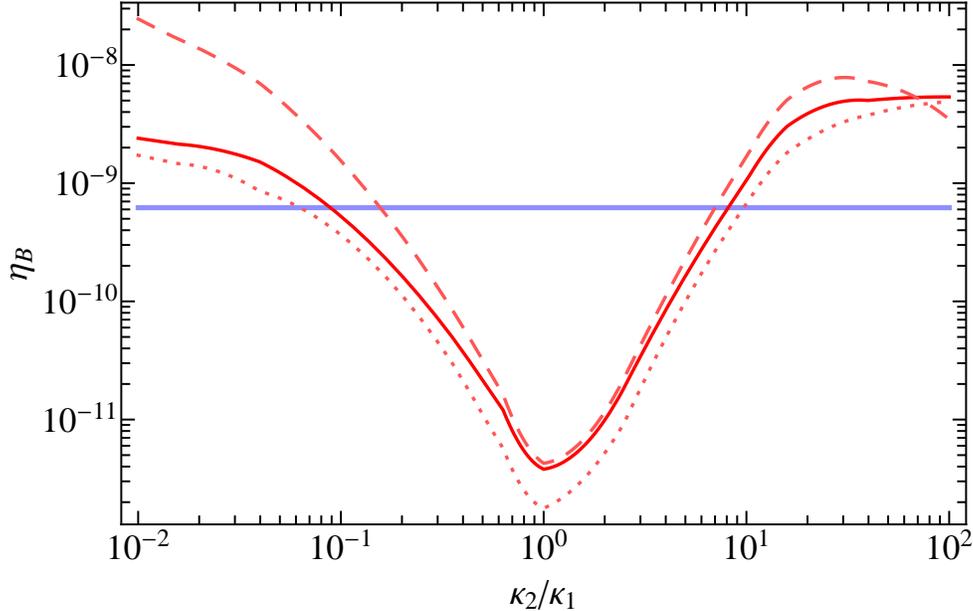}
\caption{\it The baryon asymmetry $\eta_B$ as a function of the ratio
  $\kappa_1/\kappa_2$ in a R$\tau$L scenario, where $\kappa_1
  \kappa_2=10^{-8}$ is fixed.  All other parameters are as in
  Figure~\ref{fig:k1_k2_RtauL_75_100}~(left panel).  The solid line
  corresponds to the full numerical solution of $\eta_B$ using the
  three-heavy-neutrino mixing formula given in~(\ref{hres3g}), the
  dashed line describes the two-heavy-neutrino mixing approximation
  with $R_{\alpha\beta}=0$, and the dotted line is obtained by
  neglecting the RIS-subtracted collision terms in the
  BE~(\ref{BEDL4}), or equivalently by setting the parameter
  $\kappa_\tau$ defined in~(\ref{kappal}) equal to 1 in the
  BE~(\ref{BEDL2}).}
\label{fig:2vs3mixing} 
\end{figure}

\begin{figure}[t]
\centering
\includegraphics[clip,width=0.49\textwidth]{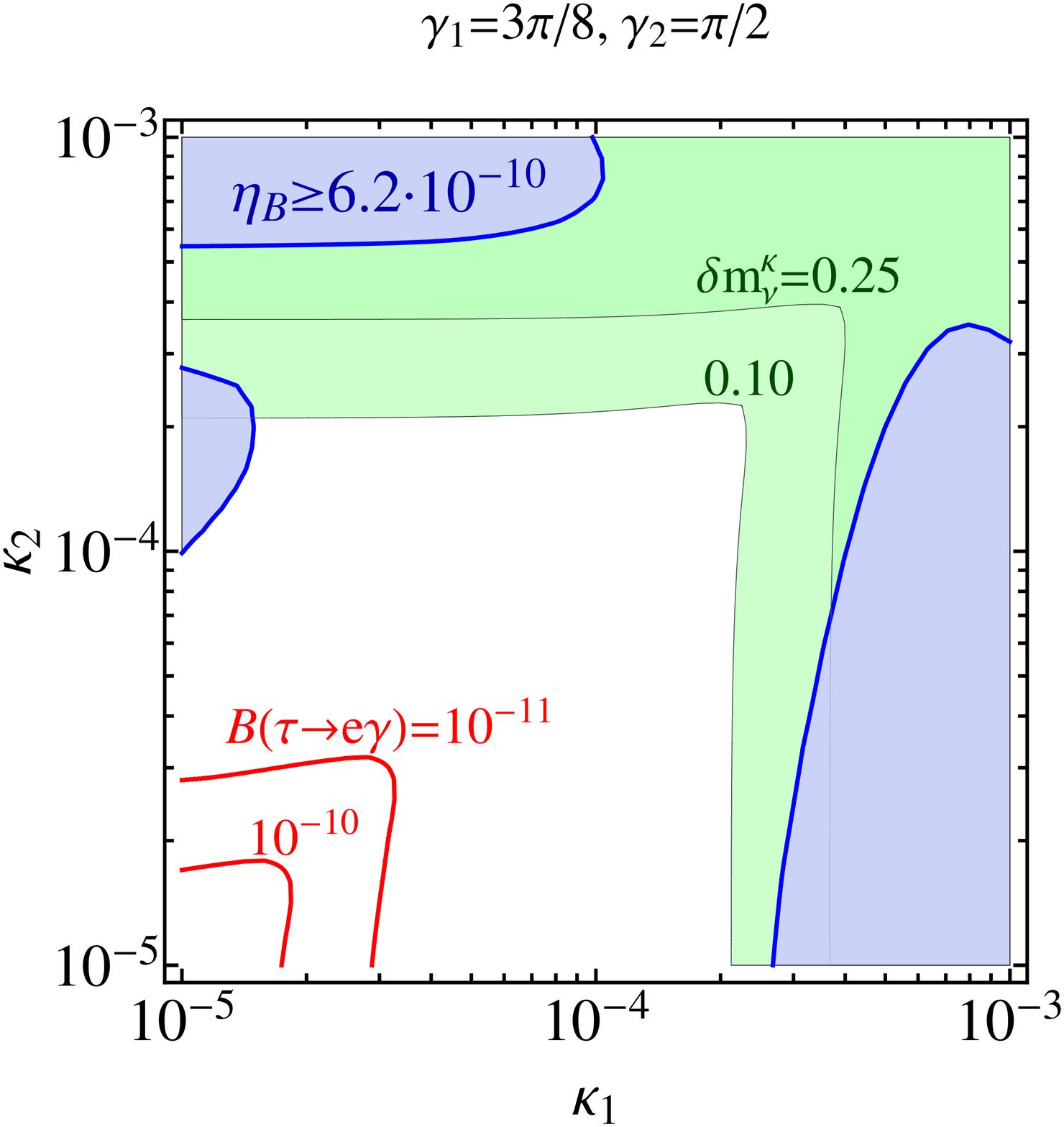}
\includegraphics[clip,width=0.49\textwidth]{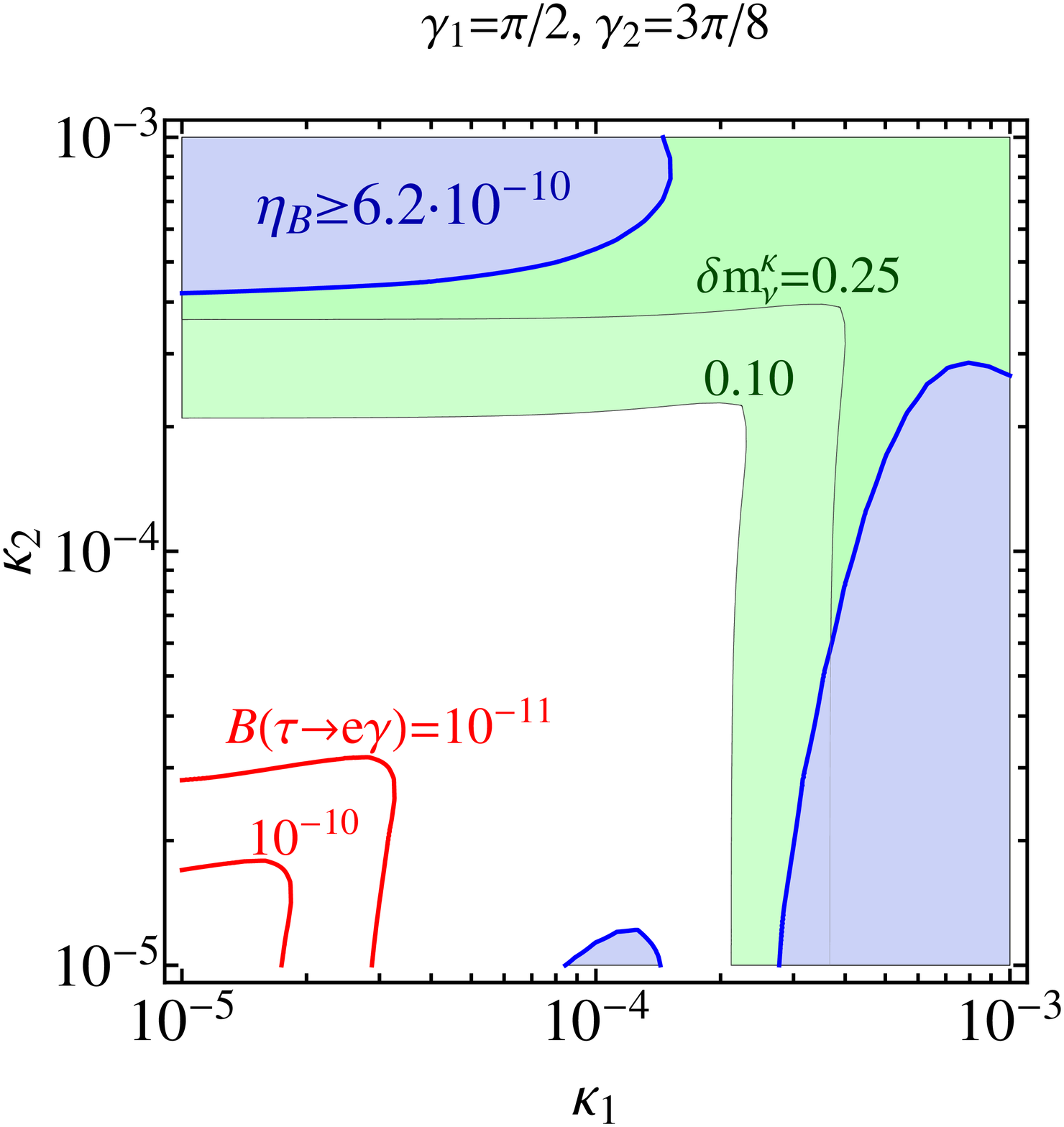}
\caption{\it  The baryon  asymmetry $\eta_B$  (blue contours)  and the
  branching ratio $B(\tau \to e\gamma)$ (red contours) as functions of
  $\kappa_1$ and  $\kappa_2$ in the R$\mu$L  model with $m_N=120$~GeV,
  where  a  normal  light  neutrino  mass spectrum  is  assumed.   The
  remaining  parameters  were  chosen as  follows:  $\gamma_1=3\pi/8$,
  $\gamma_2=\pi/2$,   $\phi_1=0$,  $\phi_2=0$,  $\text{Re}(a)>0$~(left
  panel);     $\gamma_1=\pi/2$,     $\gamma_2=3\pi/8$,     $\phi_1=0$,
  $\phi_2=\pi$,   $\text{Re}(a)<0$~(right    panel).    The   neutrino
  oscillation  parameters  are set  at  their  best  fit values,  with
  $\sin^2\theta_{13}=0.033$  at its $2\sigma$  upper limit.   The blue
  shaded regions denote the parameter space where the baryon asymmetry
  is larger than  the observational value $\eta_B=6.2\times 10^{-10}$.
  The     green     shaded     areas     labeled     as     `$\,\delta
  \mathrm{m}_\nu^\kappa=0.25$'  and `$\,0.10$' indicate  the parameter
  space  where the  inversion  of the  light-neutrino  mass matrix  is
  violated at the 25\% and 10\% levels, respectively.}
\label{fig:k1_k2_RmuL_75_100} 
\end{figure}

\begin{figure}[t]
\centering
\includegraphics[clip,width=0.49\textwidth]{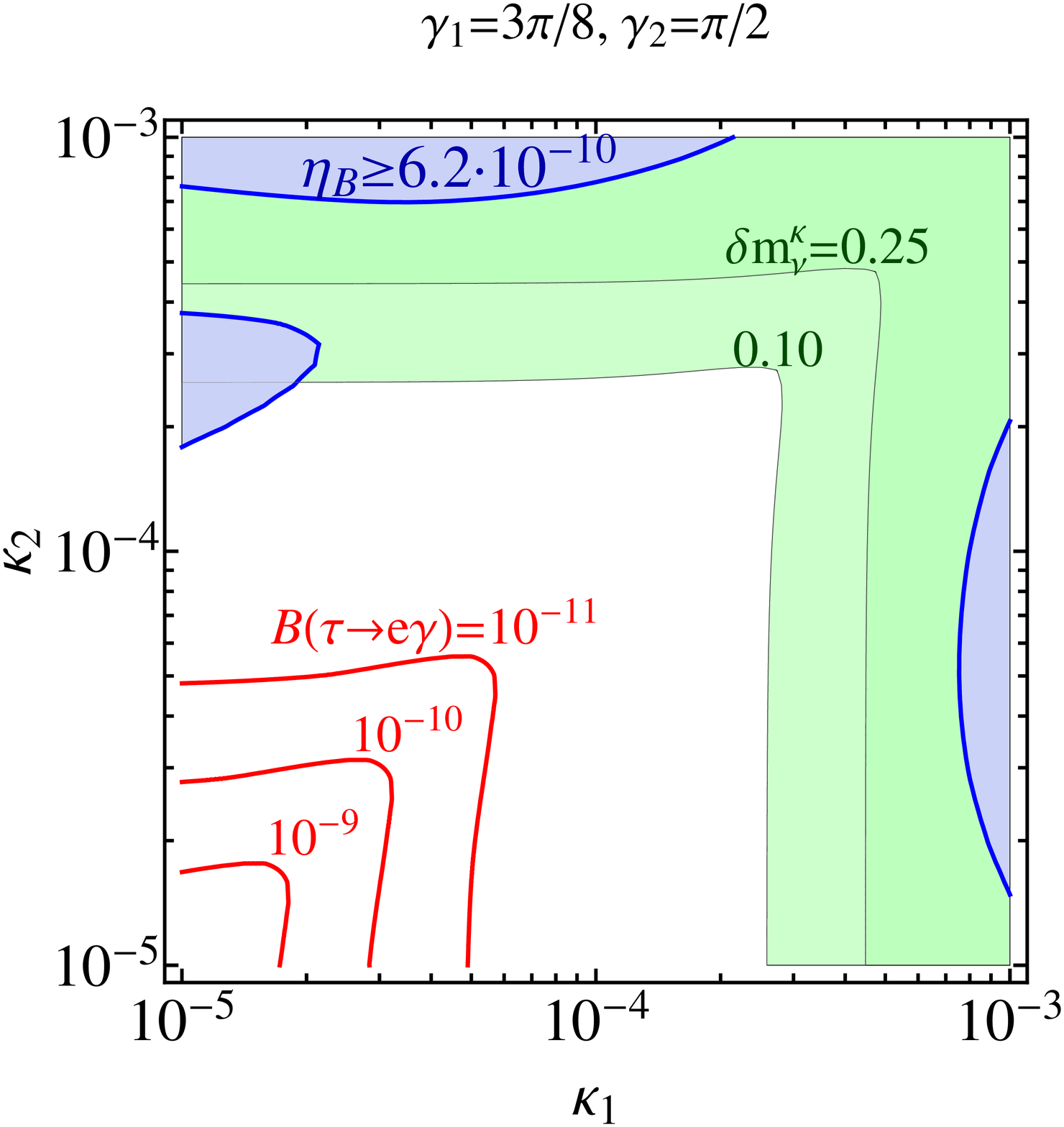}
\includegraphics[clip,width=0.49\textwidth]{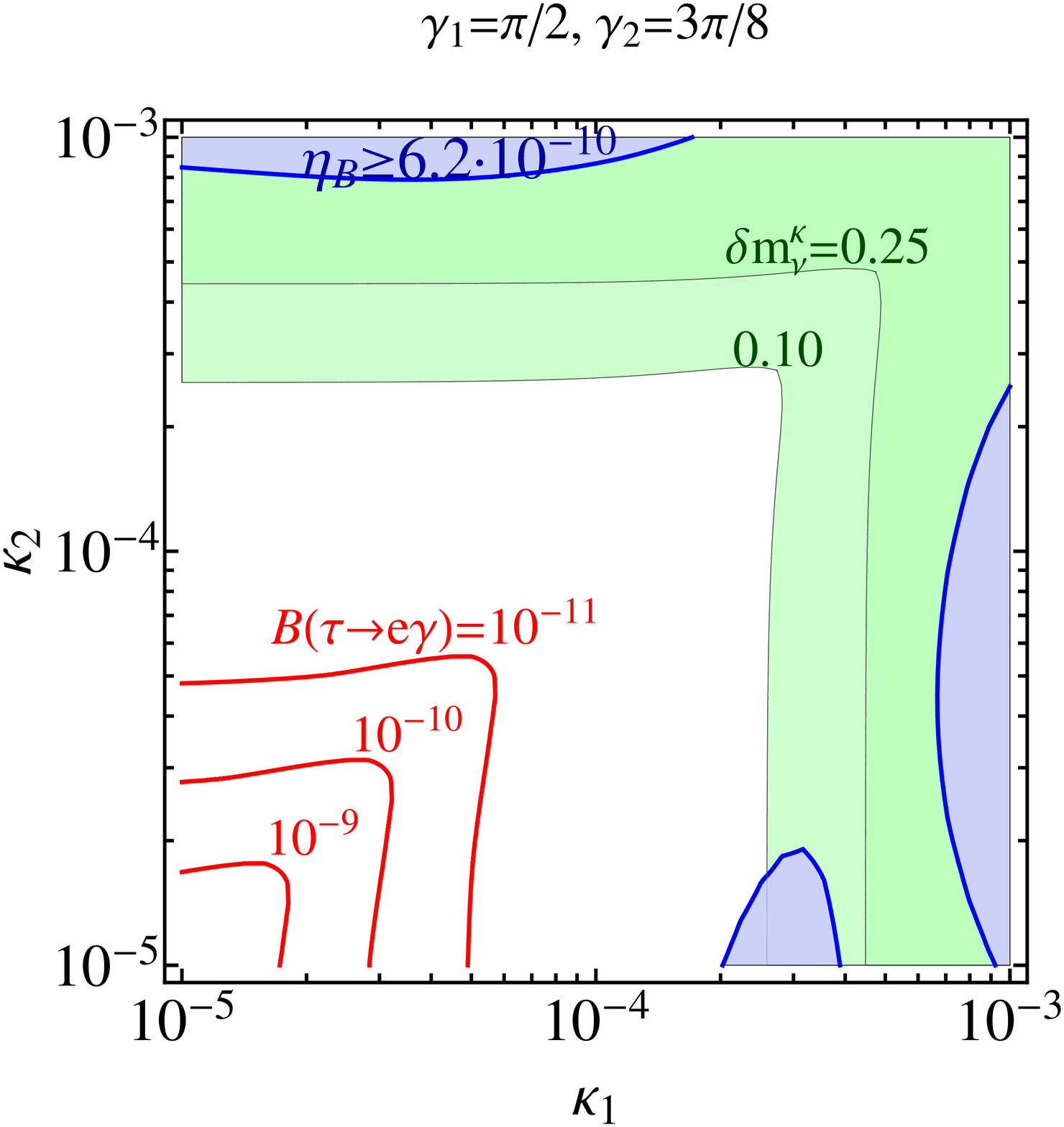}
\caption{\it The same as in Figure~\ref{fig:k1_k2_RmuL_75_100}, but
  for an inverted light neutrino mass spectrum and the following
  choice of parameters: $\gamma_1=3\pi/8$, $\gamma_2=\pi/2$,
  $\phi_1=0$, $\phi_2=0$, $\text{Re}(a)<0$ (left panel);
  $\gamma_1=\pi/2$, $\gamma_2=3\pi/8$, $\phi_1=0$, $\phi_2=0$,
  $\text{Re}(a)>0$ (right panel).}
\label{fig:k1_k2_RmuL_75_100_inv} 
\end{figure}

\begin{figure}[t]
\centering
\includegraphics[clip,width=0.49\textwidth]{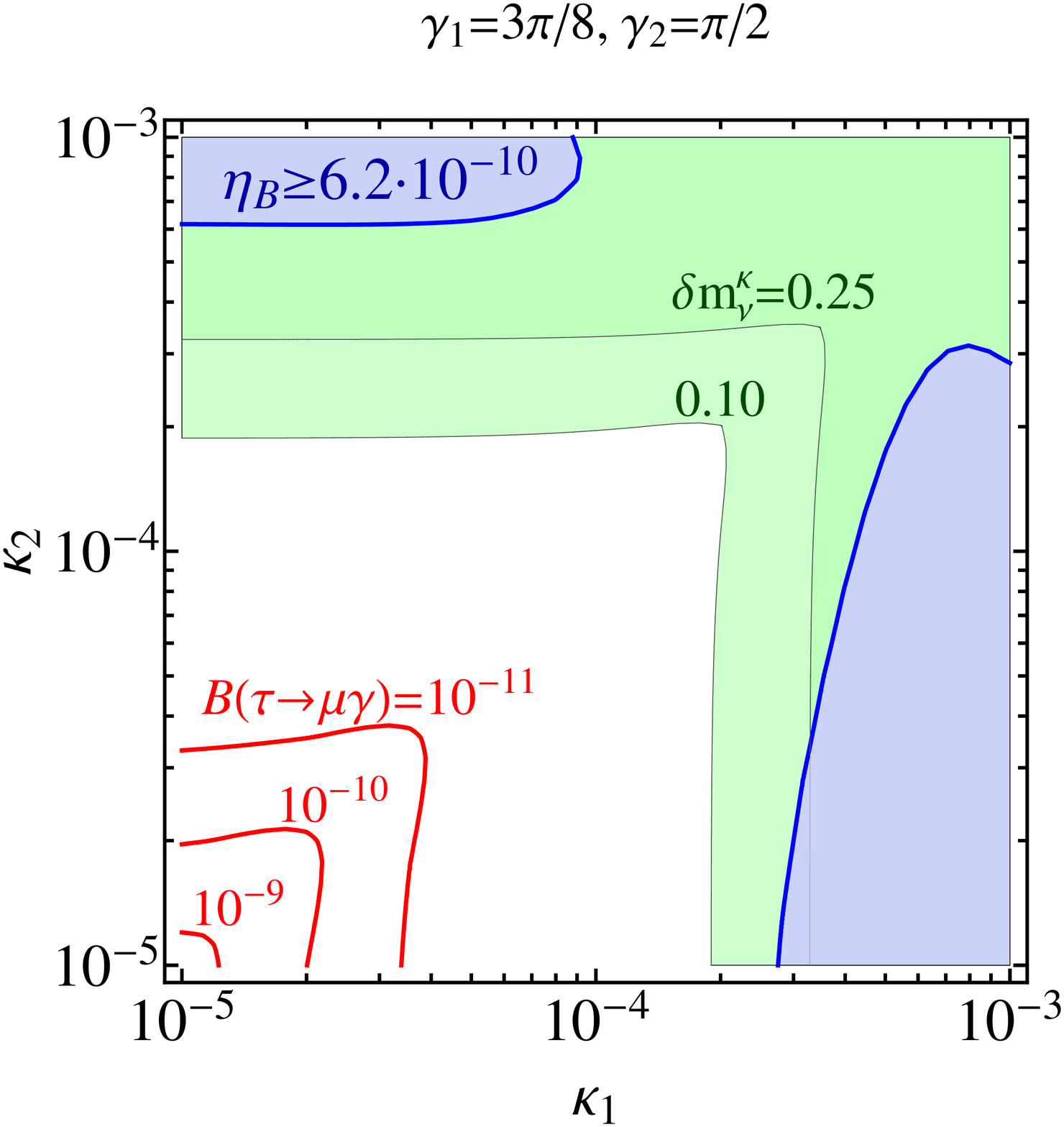}
\includegraphics[clip,width=0.49\textwidth]{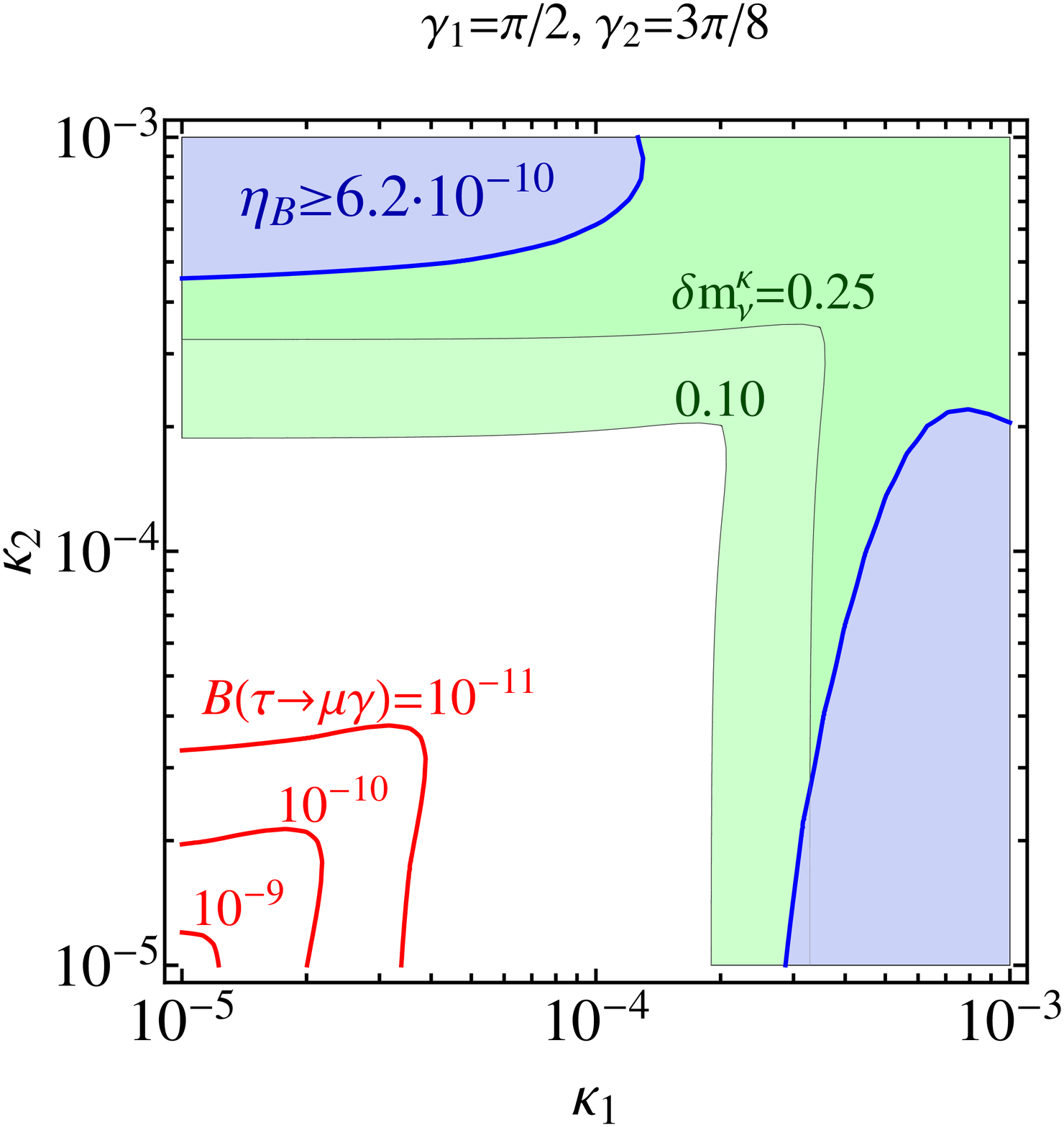}
\caption{\it  The baryon  asymmetry $\eta_B$~(blue  contours)  and the
  branching ratio $B(\tau \to \mu \gamma)$~(red contours) as functions
  of $\kappa_1$ and $\kappa_2$  in the R$e$L model with $m_N=120$~GeV,
  considering a  normal light  neutrino mass spectrum.   The following
  choice    of    the    remaining   parameters:    $\gamma_1=3\pi/8$,
  $\gamma_2=\pi/2$,  $\phi_1=\pi$,  $\phi_2=0$, $\text{Re}(a)>0$~(left
  panel);     $\gamma_1=\pi/2$,     $\gamma_2=3\pi/8$,     $\phi_1=0$,
  $\phi_2=\pi$,   $\text{Re}(a)>0$~(right    panel).    The   neutrino
  oscillation parameters are chosen at  their best fit values but with
  $\sin^2\theta_{13}=0.033$  at its $2\sigma$  upper limit.   The blue
  shaded regions denote the parameter space where the baryon asymmetry
  is  larger  than $\eta^{\rm  obs}_B=6.2\times  10^{-10}$. The  green
  shaded  areas labeled  as `$\,  \delta{\rm  m}_\nu^\kappa=0.25$' and
  `$\, 0.10$' indicate the parameter  space where the inversion of the
  light-neutrino mass matrix is violated  at the 25\% and 10\% levels,
  respectively.}
\label{fig:k1_k2_ReL_75_100} 
\end{figure}

\begin{figure}[t]
\centering
\includegraphics[clip,width=0.49\textwidth]{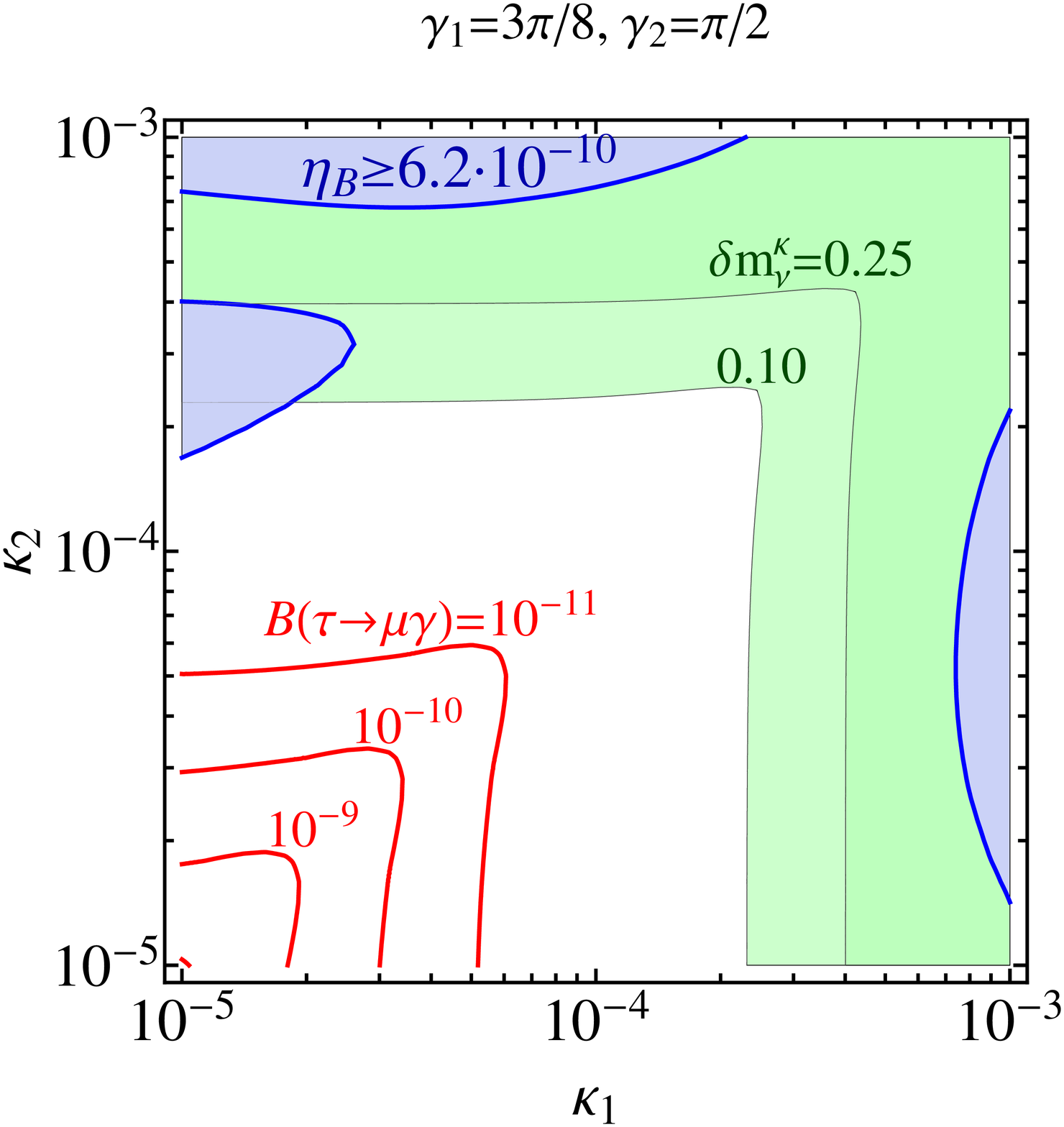}
\includegraphics[clip,width=0.49\textwidth]{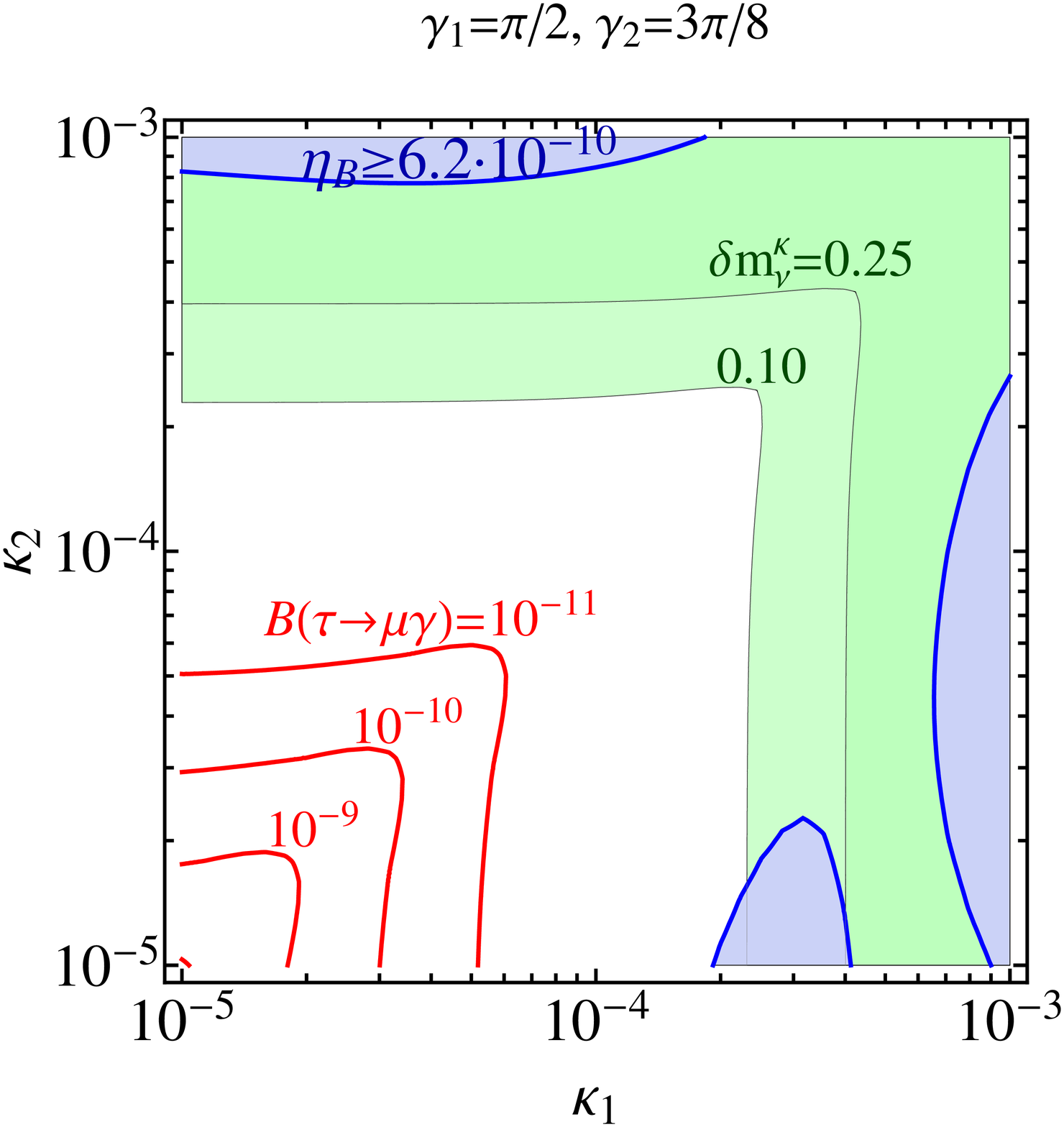}
\caption{\it The same as in Figure~\ref{fig:k1_k2_ReL_75_100}, but for
  an inverted light neutrino mass spectrum and the following choice of
  parameters: $\gamma_1=3\pi/8$, $\gamma_2=\pi/2$, $\phi_1=0$,
  $\phi_2=\pi$, $\text{Re}(a)<0$~(left panel); $\gamma_1=\pi/2$,
  $\gamma_2=3\pi/8$, $\phi_1=0$, $\phi_2=0$, $\text{Re}(a)>0$~(right
  panel).}
\label{fig:k1_k2_ReL_75_100_inv} 
\end{figure}

For a  given light  neutrino mass matrix  ${\bf m}^\nu$,  the solution
obtained for  $a$ and $b$ is unique  up to a common  sign factor. This
sign   degeneracy  could   be  eliminated,   only  if   the   sign  of
$\text{Re}(a)$ were  known.  There  is also a  similar freedom  for an
overall sign in the $\epsilon_{e,\mu,\tau}$ parameters, but this turns
out to  be irrelevant, since  it applies to  a complete column  in the
neutrino  Yukawa matrix  ${\bf h}^\nu$  and can  be rotated  away.  In
addition to the light neutrino  masses and mixings, the light neutrino
mass matrix may also contain the CP-violating Dirac phase $\delta$ and
the  Majorana phases  $\phi_{1,2}$.  As  our ansatz  for  the neutrino
Yukawa matrix  ${\bf h}^\nu$ uses maximal  CP phases for  $a$ and $b$,
the  inclusion of  the light  neutrino CP  phases is  not  expected to
increase  or  change  significantly  our predictions  for  the  baryon
asymmetry $\eta_B$.   We therefore consider only the  extreme cases of
CP parities,  $\delta,\phi_{1,2}=0,\pi$, as natural  choices which can
help us to reduce the  dimensionality of the parameter space.  As well
as   $\sin^2\theta_{12}$,   $\sin^2\theta_{23}$,  $\sin^2\theta_{13}$,
$\Delta   m^2_{12}$,  $|\Delta  m^2_{13}|$,   the  sign   of  $(\Delta
m^2_{13})$,  $\delta$   and  $\phi_{1,2}$,  we  also   have  the  free
parameters  $\kappa_{1,2}$, $\gamma_{1,2}$,  the sign  of  Re$(a)$ and
$m_N$ in  our R$\ell$L  models.  Unless otherwise  stated, we  use the
best fit values of~\cite{JVdata} for the measured neutrino oscillation
parameters.

To   start  with,   we  show   in  Figure~\ref{fig:k1_k2_RtauL_75_100}
numerical  estimates of  the  baryon asymmetry  $\eta_B$  and the  LFV
observables  $B(\mu  \to  e\gamma)$  and  $B_{\mu  e}({}^{48}_{22}{\rm
  Ti})$, as functions of the Yukawa-coupling parameters $\kappa_1$ and
$\kappa_2$,  in  a R$\tau$L  model  with  $m_N=120$~GeV  and a  normal
hierarchical light  neutrino mass spectrum. The  neutrino mixing angle
$\sin^2\theta_{13}$  is  set   to  its  upper  experimental  $2\sigma$
limit~\cite{JVdata}:    $\sin^2\theta_{13}=0.033$.     The   remaining
theoretical  parameters are  chosen  to maximise  the overlap  between
successful generation  of the required  baryon asymmetry and  high LFV
rates.  For definiteness,  we set $\gamma_1=3\pi/8$, $\gamma_2=1/2\pi$
(left panel)  and $\gamma_1=1/2\pi$, $\gamma_2=3\pi/8$  (right panel).
The blue shaded areas denote the parameter space where the numerically
predicted baryon  asymmetry~$\eta_B$ is larger  than the observational
value $\eta^{\rm obs}_B =  6.2\times 10^{-10}$.  These areas should be
regarded  as  representing  regions  of  viable  parameter  space  for
successful  leptogenesis, given  the  freedom of  re-adjusting the  CP
phases $\gamma_{1,2}$ of the Yukawa couplings~$\kappa_{1,2}$.

The   parameters   $\kappa_{1,2}$   are   varied  within   the   range
$10^{-5}$--$10^{-3}$; smaller  values of $\kappa_{1,2}$  would lead to
too large  values for $a,b>0.1$, whereas  larger $\kappa_{1,2}$ values
would  violate   the  assumed  approximation   $\kappa_{1,2}\ll  a,b$,
required to  invert the  seesaw formula.  The  degree of  violation of
this  approximation  is  indicated   by  the  green  shaded  areas  in
Figure~\ref{fig:k1_k2_RtauL_75_100}.  These  green shaded areas denote
the parameter space, in which the quantity
\begin{equation}        
\delta {\rm m}_\nu^\kappa\ \equiv\ \frac{\Delta m_N}{m_N} \left(
\frac{\text{max}(\kappa_1,\kappa_2)}
     {\text{min}(|\epsilon_e|,|\epsilon_\mu|,|\epsilon_\tau|)} \right)^2
\end{equation}
is bigger than 0.25 (dark green) and 0.10 (light green), respectively.
The quantity $\delta {\rm  m}_\nu^\kappa$ is a measure that quantifies
the  accuracy  of  our   analytic  approximation  for  neglecting  the
contribution of  the neutrino  Yukawa couplings $\kappa_{1,2}$  in the
light  neutrino  mass   matrix  ${\bf  m}^\nu$  [cf.~(\ref{mnutree})].
Hence, the values of $\delta  {\rm m}_\nu^\kappa = 0.10$ and 0.25 mean
that  the inversion of  the  light neutrino  mass matrix  is
accurate at the 10\% and  25\% level, respectively, due to the assumed
absence of  the $\kappa_{1,2}$ terms.  It~should be  stressed that the
corresponding  parameter space  is  not ruled  out;  larger values  of
$\kappa_{1,2}\gtrsim  5\times  10^{-4}$   would  require  a  numerical
approach  to  invert the  light  neutrino  mass matrix~${\bf  m}^\nu$,
beyond our analytic approximation.

The    specific   choice    for   the    phases    $\gamma_{1,2}$   in
Figure~\ref{fig:k1_k2_RtauL_75_100}    approximately   maximises   the
numerically   predicted  baryon  asymmetry~$\eta_B$   compatible  with
testable   LFV   decay   rates.    Specifically,  the   areas   around
$(\kappa_1,\kappa_2)\approx   (10^{-5},10^{-4})$   (left  panel)   and
$(10^{-4},10^{-5})$  (right panel),  where $B(\mu\to  e\gamma) \approx
10^{-12}$  can   be  achieved,  are  only   present  for  sufficiently
asymmetric  values $\gamma_1  \neq  \gamma_2$, and  only for  specific
choices for  the remaining discrete parameters in  the R$\tau$L model.
As  a  consequence,  the  requirement  of  both  a  successful generation of the baryon
asymmetry~$\eta_B$ and  potentially observable LFV rates  for $\mu \to
e$ transitions  puts severe constraints on the  model parameter space.
We note  that the dependence  of the baryon asymmetry~$\eta_B$  on the
right-handed  neutrinos mass  scale $m_N$  is weak  in  the physically
interesting region  of $m_N =$~100--500~GeV.  Finally,  the LFV $\tau$
decays  are extremely  suppressed with  $B(\tau \to  l_2  \gamma) \sim
10^{-17}$  ($l_2=e,\mu$),  and  so  remain  far beyond  the  realm  of
detection.

In   Figure~\ref{fig:k1_k2_RtauL_75_100_inv}   we  display   numerical
estimates of $\eta_B$ and the LFV observables $B(\mu \to e\gamma)$ and
$B_{\mu  e}({}^{48}_{22}{\rm  Ti})$, as  functions  of $\kappa_1$  and
$\kappa_2$,  in a R$\tau$L  model with  $m_N=120$~GeV and  an inverted
hierarchical  light neutrino mass  spectrum, characterised  by $\Delta
m^2_{13}<0$.    As  before,   we   set  the   neutrino  mixing   angle
$\sin^2\theta_{13}$    to    its    upper    experimental    $2\sigma$
limit~\cite{JVdata}: $\sin^2\theta_{13}=0.033$.   As can be  seen from
Figure~\ref{fig:k1_k2_RtauL_75_100_inv}, the  regions of the parameter
space that yield successful baryon asymmetry $\eta_B$ are smaller than
those found  in the  normal hierarchical case  for the  light neutrino
mass  spectrum.  If  such a  scenario  gets realized  in nature,  then
successful leptogenesis implies  rates for $B_{\mu e}({}^{48}_{22}{\rm
Ti}) \stackrel{<}{{}_\sim} 10^{-16}$, which  can still be within reach
of the projected PRISM experiment~\cite{PRISM}.

The higher  rates for the LFV  $\mu \to e$ transitions  in a R$\tau$L
model with  an inverted hierarchical light-neutrino  mass spectrum may
be  attributed  to  the  fact  that  the  squared  parameter~$a^2$  is
proportional  to the  not  yet well  determined neutrino-mixing  angle
$\sin^2\theta_{13}$.   In  detail,  the  branching  ratio  $B(\mu  \to
e\gamma)$ is  enhanced in the  case of an inverted  hierarchical light
neutrino mass spectrum ($\Delta  m^2_{13}<0$).  Instead, in the normal
hierarchical light-neutrino scenario  with $\Delta m^2_{13}>0$, $B(\mu
\to e\gamma)$  is essentially independent  of $\sin\theta_{13}$.  This
is   demonstrated  in   Figure~\ref{fig:theta13},   showing  $B(\mu\to
e\gamma)$    as    a    function   of    $\sin^2\theta_{13}$,    where
$\kappa_{1}=10^{-5}$ and $\kappa_2=10^{-4}$.   In the same figure, the
vertical solid and dashed lines  denote the current best fit value and
the   expected  sensitivity  of   future  experiments   for  measuring
$\sin^2\theta_{13}$,  respectively.   The  branching  ratio  $B(\mu\to
e\gamma)$  depends  linearly on  $\sin^2\theta_{13}$  in the  inverted
hierarchical   light-neutrino  scenario  where   $\Delta  m^2_{13}<0$.
Unlike  in the  normal  light-neutrino scenario,  the predicted  value
for~$\eta_B$ in the R$\tau$L model with an inverted hierarchical light
neutrino mass spectrum falls short of explaining the BAU by two orders
of magnitude,  for a potentially observable branching  ratio of $B(\mu
\to e\gamma) \sim 10^{-13}$.

In  Figure~\ref{fig:2vs3mixing}   we  illustrate  the   importance  of
including the  full three-heavy-neutrino mixing in  the calculation of
the  baryon  asymmetry.  We  present  a  comparison  between the  full
calculation   based   on   using   the  effective   Yukawa   couplings
$\overline{\bf h}^\nu_{l \alpha}$ given in~(\ref{hres3g}) (solid line)
and its two-heavy-neutrino mixing approximation (dashed line) where we
set  $R_{\alpha\beta}=0$.   Note  that the  two-heavy-neutrino  mixing
approximation can differ from the  full calculation by up to one order
of  magnitude.   In  addition, Figure~\ref{fig:2vs3mixing}  shows  the
baryon  asymmetry~$\eta_B$ calculated  by omitting  the RIS-subtracted
collision terms in the BE~(\ref{BEDL4}) (dotted line), or equivalently
by taking the  parameter $\kappa_\tau$ defined in~(\ref{kappal}) equal
to 1  in the BE~(\ref{BEDL2}).   Such a simplification may  reduce the
predicted values for $\eta_B$ by as much as 60\%.

Figures~\ref{fig:k1_k2_RmuL_75_100}                                 and
\ref{fig:k1_k2_RmuL_75_100_inv}  present  numerical  estimates, for  a
R$\mu$L scenario with normal and inverted light neutrino mass spectra,
respectively.  We  see that  the baryon asymmetry~$\eta_B$  exhibits a
similar  dependence on $\kappa_1$  and $\kappa_2$  as in  the R$\tau$L
scenario.  Due  to the large $e$- and  $\tau$-Yukawa couplings present
in the  R$\mu$L model,  the largest  LFV rate can  be observed  in the
$\tau \to  e$ transitions, e.g.~in the LFV  process $\tau\to e\gamma$,
with  $B  (\tau  \to  e\gamma)  \sim  10^{-10}$.   Since  the  current
experimental  sensitivity to  this process  is  $B_\text{exp}(\tau \to
e\gamma) \approx  10^{-7}$ which is  not expected to increase  by more
than one  order of  magnitude in the  foreseeable future,  it~would be
difficult  to  probe  the  parameter  space of  the  R$\mu$L  scenario
compatible with  observable BAU.   On the other  hand, the  $\mu\to e$
transitions  in  the R$\mu$L  model,  although  being proportional  to
$\text{max}(\kappa_1,\kappa_2)^2   a^2$  and   so  smaller   than  the
predictions obtained  in R$\tau$L scenario, are  still sizeable enough
to    produce   a    $\mu\to   e$    conversion   rate    of   $B_{\mu
e}(^{48}_{22}\text{Ti})   \approx  2\times   10^{-17}$   (for  $\Delta
m_{13}^2>0$) and $5\times 10^{-16}$ (for $\Delta m_{13}^2<0$).  As the
parameters  $a$ and  $b$ are  approximately inversely  proportional to
$\kappa_{1,2}$,  these  values  are  largely  independent  of
$\kappa_{1,2}$ and apply  to the whole $(\kappa_1,\kappa_2)$ parameter
plane,   as   depicted   in  Figures~\ref{fig:k1_k2_RmuL_75_100}   and
\ref{fig:k1_k2_RmuL_75_100_inv}. Consequently,  this scenario could be
probed  at  a  future  $\mu   \to  e$  conversion  experiment  with  a
sensitivity    of   $\sim    10^{-16}$~(COMET,    mu2e),   or    $\sim
10^{-18}$~(PRISM)~\cite{PRISM}.

Finally,             Figures~\ref{fig:k1_k2_ReL_75_100}            and
\ref{fig:k1_k2_ReL_75_100_inv}  display  numerical  estimates,  for  a
R$e$L scenario with normal  and inverted light neutrino mass spectra,
respectively.  Our  results are quite  analogous to the  R$\mu$L case.
Correspondingly,  the largest  LFV rate  is obtained  for  the process
$\tau   \to   \mu\gamma$.     However,   successful   R$e$L   requires
$B(\tau\to\mu\gamma) \sim 10^{-14}$, which  is far beyond the reach of
the next generation  experiments. In analogy to the  R$\mu$L case, the
rates for  coherent $\mu\to  e$ conversion in  nuclei are found  to be
sizeable.   Specifically,  we  obtain $B_{\mu  e}(^{48}_{22}\text{Ti})
\approx  3\times  10^{-17}$  (for  $\Delta m_{13}^2>0$)  and  $7\times
10^{-16}$ (for $\Delta m_{13}^2<0$). Interestingly enough, these rates
are well within reach of the proposed PRISM experiment~\cite{PRISM}.

\section{Conclusions}\label{sec:Conclusion}

We have analyzed minimal low-scale  seesaw scenarios of resonant leptogenesis and studied
their potential  implications for observables of charged  LFV, such as
$\mu  \to e\gamma$  and $\mu  \to e$  conversion in  nuclei.   We have
considered  three physically interesting  flavour realisations  of resonant leptogenesis,
where the  observed BAU originates from an  individual $\tau$-, $\mu$-
or $e$-lepton-number asymmetry which  gets resonantly enhanced via the
out-of-equilibrium   decays  of   nearly  degenerate   heavy  Majorana
neutrinos.

By means  of approximate lepton-flavour symmetries, we  have been able
to  construct viable and  natural models  of R$\ell$L  compatible with
universal  right-handed neutrino masses  at the  GUT scale,  where the
required heavy-neutrino mass splittings  are generated via RG effects.
Particular  attention has  been  paid that  the effective  resummation
method introduced in~\cite{APRD} and used  in our study to compute the
resonantly   enhanced  lepton  asymmetries   respects  the   Nanopoulos-Weinberg  no-go
theorem~\cite{NW}   in  the  $L$-conserving   limit  of   the  theory.
Specifically,   we  have   checked  that   the   leptonic  asymmetries
$\delta_{\alpha  l}$ given in~(\ref{deltaN})  and~(\ref{dCPla}) vanish
in all  parametrically possible $L$-conserving limits  of the R$\ell$L
scenarios.   In agreement  with  earlier studies~\cite{APtau,PU2},  we
find that  at least  three heavy Majorana  neutrinos are  required, in
order   to   potentially  have   both   successful  leptogenesis   and
experimentally  testable  rates for  LFV  processes,  such as  $\mu\to
e\gamma$ and $\mu \to e$ conversion in nuclei.

We have found that the  heavy Majorana neutrinos in R$\ell$L scenarios
can  be as  light as  100~GeV, whilst  their couplings  to two  of the
charged leptons may  be large so as to lead to  LFV effects that could
be tested by the MEG and the COMET/PRISM experiments.  Specifically, in
the R$\tau$L model with a  normal light neutrino mass hierarchy, there
is a  sizeable model parameter space with  successful leptogenesis and
large  LFV process  rates, with  $B(\mu\to  e\gamma)\approx 10^{-12}$.
This prediction is largely  independent of $\sin^2\theta_{13}$ and the
other light  neutrino oscillation parameters.   On the other  hand, in
the  R$\tau$L  model  with  inversely  hierarchical  light  neutrinos,
$B(\mu\to e\gamma)$  is linearly proportional  to $\sin^2\theta_{13}$,
and can  be enhanced by more  than one order of  magnitude compared to
the normal  hierarchy case for $\sin^2\theta_{13}$ close  to its upper
experimental  $2\sigma$ limit.   Unfortunately,  the generated  baryon
asymmetry~$\eta_B$  is suppressed in  this scenario,  and to  test the
viable parameter  space for  successful leptogenesis would  require an
experiment for $\mu \to e$  conversion in nuclei which is sensitive to
$B_{\mu e}\approx  10^{-17}$--$10^{-16}$.  This feature  is also quite
generic for  the case of the  R$\mu$L and R$e$L  models, where $\mu
\to   e$   flavour   transitions   are   suppressed   by   the   small
$\tau$-Yukawa couplings~$\kappa_{1,2}$.  In all R$\ell$L  models, charged LFV in the
$\tau$-lepton sector  turns out to be  at least 6  orders of magnitude
beyond  the   current  experimental  sensitivity,   as  the  predicted
branching   ratios   are   $B(\tau   \to   e\gamma   ,   \mu   \gamma)
\stackrel{<}{{}_\sim}  10^{-14}$  in  parameter regions  required  for
successful leptogenesis.

Further  studies  will  be  needed   to  analyze  the  full  range  of
theoretical, phenomeno\-logical  and cosmological implications  of the
three different universal models  of R$e$L, R$\mu$L and R$\tau$L.  For
instance,  the  consideration  of thermal  effects~\cite{Thermal}  may
provide  a  significant  improvement  on  the  standard  framework  of
classical BEs adopted in  the present analysis. An equally significant
issue is whether our minimal  RL models can account for the well-known
problem of cold Dark Matter (CDM) in the Universe. An obvious solution
would   be   to   consider   supersymmetric   versions   of   R$\ell$L
scenarios~\cite{GPP,SusyL}  and  study  the  relic  abundance  of  the
lightest stable  SUSY particle, which could be  a thermal right-handed
sneutrino~\cite{DP}.  Alternatively,  one may consider scale-invariant
extensions  of  the  SM with  right-handed  neutrinos~\cite{FKMV,ANP},
which are  minimally augmented with  one complex singlet  scalar field
whose one-loop  induced VEV  can naturally explain  the origin  of the
electroweak-scale mass  of the heavy Majorana neutrinos.   It has been
shown recently~\cite{ANP} that a minimal ${\bf Z}_4$-symmetric variant
of these models can stay perturbative  up to the Planck scale, as well as
provide   a   CDM  candidate   through   the  so-called   Higgs-portal
mechanism~\cite{HiggsPortal}.   It is  therefore rather  motivating to
perform  a  dedicated analysis  of  observing electroweak-scale  heavy
Majorana  neutrinos  within  the  specific  context  of  the  R$\ell$L
scenarios   studied    here,   through   their    possible   like-sign
dilepton~\cite{AZPC,LHCN}   and/or   trilepton~\cite{delAguila:2008cj}
signatures at the LHC or at other future high-energy colliders.

\bigskip
\subsection*{Acknowledgements}

This work is supported in part by the STFC research grant: PP/D000157/1.

\end{document}